\documentclass[12pt, draftclsnofoot, onecolumn]{IEEEtran}
\usepackage[T1]{fontenc}
\usepackage[latin9]{inputenc}
\usepackage{verbatim}
\usepackage{cite}
\usepackage{calc}
\usepackage{amsmath}
\usepackage{amssymb}
\usepackage{graphicx}
\makeatletter
\usepackage[caption=false,font=footnotesize]{subfig}
\@ifundefined{showcaptionsetup}{}{%
	\PassOptionsToPackage{caption=false}{subfig}}
\usepackage{subfig}
\makeatother
\begin{document}

\title{Multi-tone Signal Optimization for Wireless Power Transfer in the
Presence of Wireless Communication Links}

\author{Boules~A. Mouris,~\IEEEmembership{Student Member,~IEEE,} Hadi~Ghauch,~\IEEEmembership{Member, IEEE,}
Ragnar~Thobaben,~\IEEEmembership{Member, IEEE,} and~B.~L.~G.~Jonsson\thanks{B. A. Mouris, R. Thobaben and B.~L.~G. Jonsson are with the School of
Electrical Engineering and Computer Science (EECS), KTH Royal Institute
of Technology, Stockholm 100 44, Sweden (e-mail:boules@kth.se).}\thanks{H. Ghauch is with the Department of COMELEC, Telecom-ParisTech, Institut
Mines-Telecom, Paris, France (e-mail: hadi.ghauch@telecom-paristech.fr).}}

\markboth{}{Boules A. Mouris \MakeLowercase{\emph{et al.}}: Multi-tone Signal
Optimization for Wireless Power Transfer in the Presence of Wireless
Communication Links}
\maketitle
%\vspace{-0.75cm}
\begin{abstract}
In this paper, we study optimization of multi-tone signals for wireless
power transfer (WPT) systems. We investigate different non-linear
energy harvesting models. Two of them are adopted to optimize the
multi-tone signal according to the channel state information available at the transmitter. We show that a second-order polynomial
curve-fitting model can be utilized to optimize the multi-tone signal
for any RF energy harvester design. We consider both single-antenna and multi-antenna WPT systems. In-band
co-existing communication links are also considered in this work by
imposing a constraint on the received power at the nearby information
receiver to prevent its RF front end from saturation. We emphasize
the importance of imposing such constraint by explaining how inter-modulation
products, due to saturation, can cause high interference at the information
receiver in the case of multi-tone signals. The multi-tone optimization
problem is formulated as a non-convex linearly constrained quadratic
program. Two globally optimal solution approaches using mixed-integer
linear programming and finite branch-and-bound techniques
are proposed to solve the problem. The achieved improvement resulting
from applying both solution methods to the multi-tone optimization
problem is highlighted through simulations and comparisons with other
solutions existing in the literature.
\end{abstract}


\IEEEpeerreviewmaketitle{}
\section{Introduction}

\IEEEPARstart{T}{he } massive increase in the number of wireless
devices in future communication systems such as Internet of Things
(IoT) and fifth-generation (5G) networks has increased the demand
to eliminate wires completely from electronics devices \protect\cite{7120022}.
Wireless power transfer (WPT) is a promising way to cross the barrier
imposed by batteries. The history of wireless power transfer dates
back to the Tesla coil more than 100 years ago. However, with recent
advancements in microwave technology and antenna design, extensive
research addressing WPT technology has been conducted over the last
decade \protect\cite{HarvestingWP,EuropeandWPTFuture}. In order to fully
power electronic devices wirelessly, two research directions have
been investigated: minimizing power consumption in electronic devices,
and improving WPT efficiency. Recently, energy-efficient wireless
sensors were proposed in \protect\cite{LowPowerAnalogJSCC}, in which analog
joint source-channel coding was exploited to compress sensor signals
allowing a power consumption as low as $90\,\mathrm{\mu W}$. Such
sensors can be fully operated by wireless power, and thus, enable
autonomous wireless sensor networks.

In the context of limiting battery usage in communication systems,
extensive research has been devoted to studying wireless powered communications
(WPC) \protect\cite{WPC} as well as simultaneous wireless information and power
transfer (SWIPT) \protect\cite{Kirkidis}. An example for WPC is \protect\cite{NunoBackscatter.},
where wireless power transfer is used to enable load modulated backscatter
communication. Resource allocation for different SWIPT techniques
in broadband systems has been discussed in \protect\cite{SWIPTErikLarss}.
The rate-energy region for different SWIPT schemes has been characterized
in \protect\cite{MIMOSWIPT}. However, these contributions assumed a linear
energy harvester model, in which the output power is a linear function
of the input power to the energy harvester. More recently, different
non-linear energy harvester models have been proposed to model the
performance of SWIPT systems more specifically \protect\cite{Sigmoid,SWIPTBruno,Chalmer}.
These models will be discussed in more details in Section III in this
paper.

The key element of a WPT system is the RF energy harvester which is
responsible for capturing the incident electromagnetic wave and converting
it efficiently into DC power. The RF-to-DC conversion efficiency of
the RF energy harvester, usually referred to as rectenna (rectifying
antenna), determines the overall efficiency of the WPT system. Designing
an efficient rectenna is a major challenge, especially at very low
input RF power levels. Many research efforts have been directed towards
improving the efficiency of rectennas by carefully designing its components
as in \protect\cite{EH2.4,AdaptiveEHLeuven,RectennaArray,HighEffRect}. In
\protect\cite{EH2.4}, a dual-polarized antenna followed by two voltage doubler
circuits were proposed to combine RF power arriving from different
polarizations. An adaptive re-configurable rectifier that maintains
its efficiency with respect to the input power level was presented
in \protect\cite{AdaptiveEHLeuven}. Different configurations of rectenna
arrays were investigated to increase the output DC power in \protect\cite{RectennaArray}.
A broadband rectenna with a highly efficient rectifier circuit was
proposed in \protect\cite{HighEffRect}. 

Lately, it has been observed experimentally that signals having high
peak-to-average power ratio (PAPR) such as chaotic, white noise and
OFDM signals \protect\cite{OptimalWFChaoticetc} can boost the RF-to-DC conversion
efficiency of RF energy harvesters. Multisine signals having high
PAPR were shown to improve the range and reliability of RFID tag systems
in \protect\cite{POWValenta}. These interesting results motivated research
on analyzing the dependency of the efficiency of the energy harvester
on its input waveform \protect\cite{POWValenta,ImpactofMultisine}. However,
these studies ignored the wireless channel in their analysis. To the
authors best knowledge, \protect\cite{WFDesign} was the first work to account
for the wireless channel and the channel state information (CSI) available
at the transmitter in addition to the non-linear behavior of the energy
harvester in designing the WPT signal. A model based on Taylor expansion
of the I-V characteristics of the diode has been adopted in \protect\cite{WFDesign}
and used for waveform design showing performance gains over linear
models when exploiting multi-tone signal excitation. In \protect\cite{WaveformZhang},
a similar work has been proposed using the I-V characteristics of
the diode utilizing integral equations instead of the Taylor expansion.
In both \protect\cite{WFDesign,WaveformZhang}, the weights (amplitudes and
phases) of the multi-tone signal resulting from maximizing the output
DC current from the energy harvester subject to a transmit power constraint
were obtained. The globally optimal phases were obtained in closed
form, although the amplitudes were not guaranteed to be globally optimal. 

In this paper, we introduce a globally optimal solution method to
the multi-tone signal optimization problem to obtain the amplitudes
of the multi-tone signal. We first use the diode non-linear model
to express the output DC current from the energy harvester. However,
in contrast to \protect\cite{WFDesign,WaveformZhang}, we formulate the output
maximization problem as linearly constrained quadratic program (LCQP),
for which we introduce two globally optimal solution methods using
either mixed-integer linear programming (MILP) or finite branch-and-bound
(BB). Secondly, we introduce a simple second-order polynomial curve-fitting
model with the same goal of designing the amplitudes and phases of
the multi-tone signals. We show by designing and simulating a 2.4
GHz energy harvester that the curve-fitting model tends to show the
same behavior for power allocation as the diode model. We also consider
the problem of multi-tone signal optimization in a MISO WPT system.
Finally, assuming in-band SWIPT, we consider preserving the information
link between co-existing communication devices by adding a constraint
on the received power at the nearby information receiver to keep it
below the saturation point of its low-noise amplifier (LNA). This
constraint was motivated by a practical experiment conducted by the
authors: A Wi-Fi channel between a router and a computer was established
(channel 7 in IEEE 802.11g). By implementing multi-tone signal WPT
with software defined radios at 2.4 GHz, a significant decrease in
the rate of information transfer (compared to single-tone WPT) has
been observed, even when the multi-tone signals are designed to not
interfere with any OFDM sub-carrier. The authors explained this by
taking into account the RF front-end of the information receiver (before
mixing and converting to base-band), hence, adding a constraint preventing
the LNA from producing high-order harmonics.

The contributions of this work can be summarized as follows: We formulate
the multi-tone optimization problem resulting from the Taylor expansion
of the I-V characteristics of the diode as a LCQP and propose two
globally optimal solution methods to obtain the amplitudes of the
multi-tone signal. Simulation results show that applying the proposed
solution methods to the multi-tone optimization problem enhances the
output of the energy harvester compared to the other power allocation
methods existing in the literature. We explain how a second-order
polynomial curve-fitting model can be utilized in the multi-tone signal
optimization problem. Our simulation results show excellent agreement
between the diode model and the curve-fitting model. The advantage
of using this curve-fitting model arises from the fact that it can
easily be adjusted in the case of multi-stage rectifiers to simplify
the formulation of the multi-tone optimization problem. We also investigate
the problem of in-band SWIPT from a practical aspect by considering
the RF front end of the information receiver. Our simulation results
confirms the importance of considering the saturation power as well
as the distance of the nearby information receiver from the power
transmitter when using multi-tone signals for WPT. 
The rest of the paper is organized as follows: Section II introduces
the system model and Section III discusses different non-linear energy
harvester models. Section IV presents the formulation and the proposed
solution methods for the multi-tone optimization problem resulting
from the adopted two non-linear energy harvesting models. The problem
of in-band SWIPT is discussed in Section V. The simulations results
are shown in Section VI. Section VII concludes the paper and discusses
possible future work.

\emph{Notation:} Matrices and vectors are represented by bold upper
and lower case letters, respectively. Non-bold symbols represent scalars.
The Euclidean norm is represented by $\left\Vert \mathbf{x}\right\Vert =\sqrt{\left|x_{1}\right|^{2}+...+\left|x_{n}\right|^{2}}$
. $\mathbb{E}\left\{ .\right\} $ denotes the expectation. We refer
to the real part of a complex quantity by $\mathcal{\Re}\mathit{\left\{ .\right\} }$.
$\mathbf{X}^{T}$ and $\mathbf{X}^{H}$ refer to the transpose and
conjugate transpose of a matrix, respectively. $\textrm{tr}\left(\mathbf{X}\right)$
represents the trace of a matrix. We let $\mathcal{CN}\left(\mu,\sigma^{2}\right)$
be a complex Gaussian random variable with mean $\mu$ and variance
$\sigma^{2}$.
\section{System Model}
\begin{figure}[tb!]
\begin{centering}
\includegraphics[width=3.5in,height=1.8in]{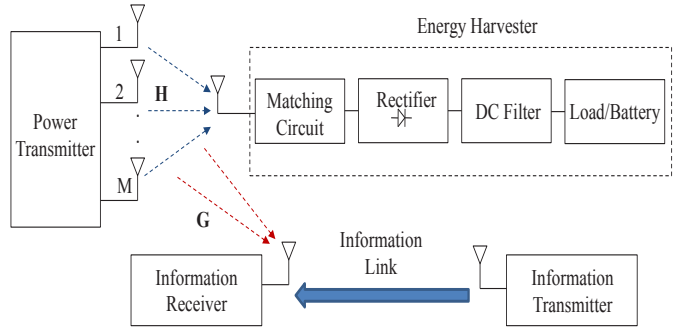}
\par\end{centering}
\centering{}\caption{Co-existing Wireless Power and Information Transfer.\label{fig:BlockDigaram}}
\vspace{-0.2in}
\end{figure}

In this paper, we study the scenario presented in Fig. \protect\ref{fig:BlockDigaram},
in which a wireless power transmitter sends power to an energy harvester
in the vicinity of an information receiver which is communicating
with an information transmitter using an OFDM-based radio access technology.
It should be noted that in our analysis, we consider interference-unaware
OFDM-based radio communication. This means that the information transmitter
and receiver are not aware of the possible interference caused by
the power transmitter. Such a scenario is prone to happen in an era
of IoT where some devices or sensors are receiving wireless power
in the vicinity of a Wi-Fi router that is communicating with smart
phones or laptop computers. The power transmitter is equipped with
$M$ antennas, while the energy harvester, the information transmitter,
and the information receiver are equipped with a single antenna for
each.

We assume that the power transmitter is transmitting $N$ tones with
frequencies $f_{n}$ such that $f_{n}=f_{o}+\left(n-1\right)\Delta f_{p}$
where $\Delta f_{p}$ is the sub-carrier spacing for the multi-tone
signal for WPT and $f_{o}$ is the carrier frequency. The transmitted
WPT signal at the $m^{th}$ antenna is given by
\begin{alignat}{1}
x_{P,m}(t) & =\sum_{n=1}^{N}s_{n,m}\cos\left(2\pi f_{n}t+\phi_{n,m}\right)=\Re\mathit{\left\{ \sum_{n=1}^{N}s_{n,m}e^{j\mathrm{2}\pi f_{n}t}e^{j\phi_{n,m}}\right\} },\label{eq:Xp}
\end{alignat}
with $s_{n,m}$ and $\phi_{n,m}$ referring to the amplitude and the
phase of the $n^{th}$ tone at antenna $m$, respectively. Let $w_{n,m}=s_{n,m}e^{j\phi_{n,m}}$,
$\mathbf{w}_{n}=[w_{n,1},w_{n,2},..,w_{n,m}]^{T}$ and $\mathbf{x_{p}}\left(t\right)=[x_{P,1}\left(t\right),x_{P,2}\left(t\right),..,x_{P,M}\left(t\right)]^{T}$.
We can write the transmitted power signal as
\begin{equation}
\mathbf{x_{p}}\left(t\right)=\mathcal{\Re}\mathit{\left\{ \sum_{n=1}^{N}\mathbf{w}_{n}e^{j\mathrm{2}\pi f_{n}t}\right\} }.\label{eq:PowTrSig}
\end{equation}
The transmitted power signal is limited by a transmit power constraint
given by $\sum_{m=1}^{M}\mathbb{E}\left\{ \left|x_{P,m}\right|\right\} ^{2}\triangleq\frac{1}{2}\left\Vert \boldsymbol{\mathbf{S}}\right\Vert ^{2}\leqslant P$,
where the matrix $\mathbf{S}=[s_{n,m}]$ contains the non-negative
amplitudes of the multi-tone signal. The factor $\frac{1}{2}$ is
due to the sinusoidal nature of the multi-tone signal. The channel
frequency response from the transmitter to the energy harvester at
$f_{n}$ can be described by the row vector
 $\mathbf{h}_{n}=[h_{n,1}e^{j\psi_{n,1}},h_{n,2}e^{j\psi_{n,2}},..,h_{n,M}e^{j\psi_{n,M}}]$
where $h_{n,m}e^{j\psi_{n,m}}$ represents the channel frequency response
at the $n^{th}$ tone of antenna $m$. The matrix $\mathbf{H}$ shown
in Fig. \protect\ref{fig:BlockDigaram} contains all the row
vectors $\mathbf{h}_{n}$ for all tones.

At the energy harvester, the received signal can be written as
\begin{equation}
y_{p}(t)=\mathcal{\Re}\mathit{\left\{ \sum_{n=1}^{N}\mathbf{h}_{n}\mathbf{w}_{n}e^{j\mathrm{2}\pi f_{n}t}\right\} }+v\left(t\right),\label{eq:RecPowerEh}
\end{equation}
where $v$ is additive white Gaussian noise with variance $\sigma^{2}$.
The noise power is assumed to be negligible compared to transmitted
power, and therefore, we eliminate the noise from further analysis
in this paper.

Assume that the frequency response of the channel from the power transmitter
to the nearby information receiver is given by $\mathbf{G}$ as shown
in Fig. \protect\ref{fig:BlockDigaram}. The rows of $\mathbf{G}$ represent
the vector channel at the $n^{th}$ tone given by $\mathbf{g}_{n}=[g_{n,1}e^{j\zeta_{n,1}},g_{n,2}e^{j\zeta_{n,2}},..,g_{n,M}e^{j\zeta_{n,M}}]$
where $g_{n,m}e^{j\zeta_{n,m}}$ represents channel frequency response
from the power transmitter at the $n^{th}$ tone and the $m^{th}$
antenna. We assume that the information transmitter is an access point
transmitting OFDM signals to the information receiver. Following this
assumption, the received signal at the nearby information receiver
due to the power transmitter is given by 
\begin{align}
y_{I}(t) & =\mathcal{\Re}\mathit{\left\{ \sum_{n=1}^{N}\mathbf{g}_{n}\mathbf{w}_{n}e^{j\mathrm{2}\pi f_{n}t}\right\} }+z_{I}\left(t\right)+v_{I}\left(t\right),\label{eq:Yirec}
\end{align}
where $v_{I}\left(t\right)$ is the noise at the receiver, $z_{I}\left(t\right)$
is the received OFDM signal due to information exchange given by 
\begin{equation}
z_{I}\left(t\right)=\sum_{l=0}^{L-1}a_{l}\cos\left(2\pi f_{l}t+\chi_{l}\right),\label{eq:XI}
\end{equation}
with $a_{l}$, $\chi_{l}$ being the amplitude and phase of the received
OFDM signal, respectively. The number of OFDM sub-carriers is $L$,
each of frequency $f_{l}$ such that $f_{l}=f_{c}+l\Delta f_{I}$
where $f_{c}$ is the carrier frequency and $\Delta f_{I}$ is the
OFDM sub-carrier spacing.

Communication systems such as Wi-Fi and certain LTE modes use time
division duplexing (TDD), which allows the power transmitter to estimate
the corresponding CSI for the link to the information receiver (i.e.,
$\mathbf{G}$) using the pilots of the information link. The channel
to the energy harvester ($\mathbf{H}$) can be estimated in a similar
way. For example, the energy harvester can have a beacon phase where
it sends a beacon signal (uplink) to the transmitter requesting energy.
Assuming reciprocal channels, the power transmitter can use the beacon
phase to acquire the CSI to the energy harvester \protect\cite{RetroBeaconCSI}.
Backscattering can be an alternative way to estimate the channel to
the energy harvester, especially, in wirelessly powered backscatter
communication systems \protect\cite{MultiAntennaBackscatterCSI}.

We aim at maximizing the output of the energy harvester and minimizing
the interference due to the multi-tone signal at the information receiver.
Several energy harvester models exist in the literature; however,
for proper modeling in the case of multi-tone excitation, only non-linear
energy harvester models should be considered. Different non-linear
energy harvester models are discussed and summarized in the following
section. 

\section{Non-linear Energy Harvester Models}
\begin{figure}[tb!]
\begin{centering}
\includegraphics[width=3in,height=2in]{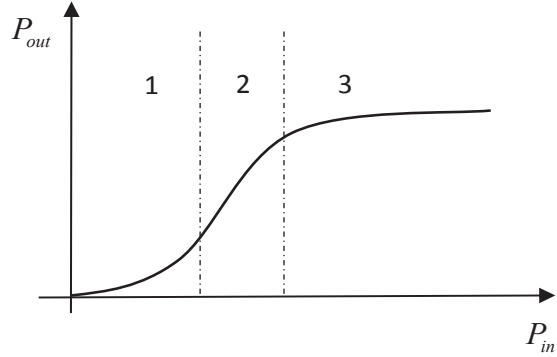}
\par\end{centering}
\caption{General input-output relationship of RF energy harvesters.\label{fig:General-input-output-relationshi}}
\vspace{-0.2in}
\end{figure}

The efficiency of RF energy harvesters is defined as the ratio between
between the output DC power of the rectifier and the input RF power
to the energy harvester. Due to the presence of non-linear elements
(e.g., diodes) in the rectifier as shown in the block diagram of the
energy harvester in Fig. \protect\ref{fig:BlockDigaram}, the efficiency of
the energy harvester is highly dependent on the input power. RF energy
harvesters are typically designed to have high conversion efficiency
at low input powers (usually few hundreds of $\mathrm{\mu W}$). The
input-output power relationship of RF energy harvesters can generally
be characterized as shown in Fig. \protect\ref{fig:General-input-output-relationshi}.
Region 1 in the figure describes the rectifier behavior in the low
input power regime. By increasing the input power to the rectifier,
the diode elements are driven to the Ohmic linear region (see Region
2) where the diode series resistance is dominant until it saturates
and the output power is approximately constant (see Region 3). For
linear operation, single-tone excitation can achieve the optimal performance
\protect\cite{SWIPTErikLarss,WFDesign}. In addition, no gains can be achieved
using multi-tone signals in Region 3 due to the rectifier saturation.
Multi-tone excitation becomes important only when the rectifier performance
is non-linear, i.e., in Region 1. Different models were proposed in
the literature to model the non-linear behavior of the rectifier.
They can be divided into two categories: the first category contains
models based on the diode non-linearity, and the second contains models
based on curve fitting. We will discuss the two different categories
in the next two subsections. 

\subsection{Rectifier Model Based on Diode Non-Linearity}

\begin{figure}[t]
\begin{centering}
\includegraphics[width=2in,height=1.3in]{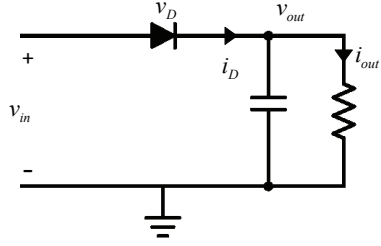}
\par\end{centering}
\caption{Single diode rectifier.\label{fig:Single-diode-rectifier}}
\vspace{-0.2in}
\end{figure}

At the energy harvester, it is assumed that the antenna is perfectly
matched to the rectifier circuit. A single-diode rectifier is shown
in Fig. \protect\ref{fig:Single-diode-rectifier} where the matching circuit
is eliminated due to the assumption of perfect matching. Single-diode
rectifiers are used in energy harvesters to minimize the power dissipated
in the rectifying components especially at very low input powers.
However, voltage doublers are sometimes preferred for full rectification
and minimizing the output ripples. Following the assumption of perfect
matching, the input voltage to the rectifier circuit can be written
from \protect\cite{WFDesign,ModelAntRect} as $v_{in}=y_{p}\left(t\right)\sqrt{R_{ant}}$,
where $R_{ant}$ is the antenna impedance. Assuming an ideal diode,
the current flowing through the diode is given by 
\begin{equation}
i_{D}=i_{s}\left(e^{\frac{v_{d}\left(t\right)}{\gamma v_{t}}}-1\right),\label{eq:Idiode}
\end{equation}
where $v_{d}\left(t\right)=v_{in}\left(t\right)-v_{out}\left(t\right)$,
$i_{s}$ is the saturation current, $\gamma$ is the non-ideality
factor, and $v_{t}$ is the diode thermal voltage. Assuming also that
the Schottky-based rectifier is in the small-signal operation region
(corresponding to Region 1 in Fig. \protect\ref{fig:General-input-output-relationshi})
and delivering constant DC output voltage (steady-state response),
it follows from the analysis presented in \protect\cite{WFDesign,MagazineOPTRectifiers,MaxDCEH}
that the output DC current of the rectifier is proportional to the
quantity
\begin{equation}
f_{DC}=k_{2}R_{ant}\mathbb{E}\left\{ y_{p}(t)^{2}\right\} +k_{4}R_{ant}^{2}\mathbb{E}\left\{ y_{p}(t)^{4}\right\} .\label{eq:fdc}
\end{equation}
This expression results from the Taylor expansion of the diode current
in (\protect\ref{eq:Idiode}), with $k_{n}$ being the expansion coefficients
which can be written as $k_{n}=\frac{i_{s}}{n!\left(\gamma v_{t}\right)^{n}}$,
and $n\in\left\{ 2,4\right\}.$ The Taylor expansion in (\protect\ref{eq:fdc})
is terminated to the $4^{th}$ harmonic only which is sufficient to
account for the diode non-linearity \protect\cite{OptimalWFChaoticetc,POWValenta}.
\subsection{Rectifier Model Based on Curve Fitting}
As an alternative to modeling the non-linear behavior of the rectifier
using the diode non-linear I-V characteristics, one can investigate
the non-linearity of the rectifier through the input vs. output power
relationship given by $P_{out}=f\left(P_{in}\right)$. Even though
the closed form expression for the input-output relation might be
unknown, it can be well approximated from simulation data using a
data-fitting tool. One of the advantages of using data-fitting models
is that they are not limited to a specific energy harvester design;
they can be adjusted to fit complicated energy harvester architectures
without the need for any assumptions. Three non-linear curve-fitting
models are summarized below: the second-order polynomial model \protect\cite{Chalmer},
the sigmoid model \protect\cite{Sigmoid}, and the rational function model
\protect\cite{RationalAlouni,RationModel1}.

\subsubsection{Second-order polynomial model}

In \protect\cite{Chalmer}, a model based on data fitting to a second-order
polynomial was proposed. Following that model, the input-output power
relation can be expressed as:
\begin{equation}
P_{out}=\beta_{1}P_{in}^{2}+\beta_{2}P_{in}+\beta_{3},\label{eq:SecondOrderP}
\end{equation}
where $\beta_{1},\,\beta_{2}$ and $\beta_{3}\in\mathcal{\mathbb{R}}$
are the parameters of the model which are obtained from curve-fitting
tools. The authors in \protect\cite{Chalmer} discussed the advantages of
the quadratic model for curve fitting by applying it to the energy
harvester presented in \protect\cite{HighEffRect} and showing that this second-order
polynomial model is concave with $\beta_{1}\leq0$, $\beta_{2}>0,$
and $\beta_{3}\leq0$. The problem of maximizing the output power
of the energy harvester is then formulated as maximizing a concave
function (i.e., a convex optimization problem) that can be solved
easily with convex optimization tools such as CVX.

\begin{figure}[tb!]
\centering{}\includegraphics[width=3.5in,height=2.5in]{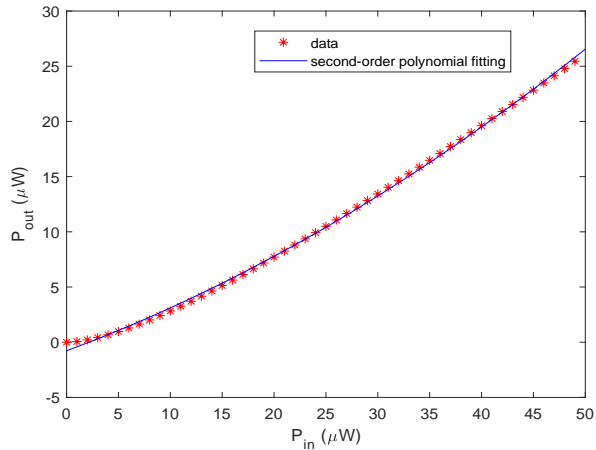}\caption{Second-order polynomial fit based on the simulations of the EH described
in Section VI. \label{fig:Curvefit}} 	\vspace{-0.2in}
\end{figure}

The above obtained choice of parameters leads to a convex optimization
problem; however, the approximated curve tend to describe the input-output
power relationship in Region 2 and Region 3 combined. If this model
is to be used in the waveform optimization problem, it should be applied
to model the behavior of the rectifier in Region 1 as shown in Fig.
\protect\ref{fig:Curvefit}. The data points in the figure are obtained from
the simulation results for the rectifier described in Section VI in
the low power region. It can be observed from Fig. \protect\ref{fig:Curvefit}
that the second order polynomial here is convex with $\beta_{1}\geq0$
and $\beta_{2}>0$. Therefore, the problem of maximizing the output
power (i.e., maximizing a convex function) in Region 1 leads to a
non-convex optimization problem which will be discussed in detail
in the next section. 
\subsubsection{Sigmoid model}

In \protect\cite{Sigmoid}, the non-linear relation between the input power
and output power for the energy harvester is modeled by the logistic
function given by
\begin{equation}
P_{out}=\frac{\frac{\pi_{3}}{1+\exp\left(-\pi_{1}\left(P_{in}-\pi_{2}\right)\right)}-\frac{\pi_{3}}{1+\exp\left(\pi_{1}\pi_{2}\right)}}{1-\frac{1}{1+\exp\left(\pi_{1}\pi_{2}\right)}},\label{eq:Sigmo}
\end{equation}
where $\pi_{1},\thinspace$$\pi_{2}$, and $\pi_{3}\in\mathcal{\mathbb{R}}$
are the parameters of the model and are obtained through data fitting.
This model provides an accurate description for the rectifier behavior
in Region 2 and Region 3 which were previously shown in Fig. \protect\ref{fig:General-input-output-relationshi}.
For proper design of the multi-tone signal, the model should consider
the behavior of the rectifier at low input power. Therefore, the sigmoid
model does not provide a complete solution to the multi-tone optimization
problem. Moreover, the sigmoid model might complicate the multi-tone
optimization problem formulation. 

\subsubsection{Rational function model}

In \protect\cite{RationModel1}, the authors investigated the possibility
to model the non-linear behavior of the rectifier through a rational
function defined by 
\begin{equation}
P_{out}=\frac{\eta_{3}P_{in}^{3}+\eta_{2}P_{in}^{2}+\eta_{1}P_{in}}{q_{3}P_{in}^{3}+q_{2}P_{in}^{2}+q_{1}P_{in}+q_{o}},\label{eq:Ration1}
\end{equation}
where $\eta_{1}$, $\eta_{2}$, $\eta_{3}$, $q_{o}$, $q_{1}$,
$q_{2}$ and $q_{3}$ are constants obtained by curve fitting. A modification
to this rational function model was introduced by \protect\cite{RationalAlouni}
to reduce the number of unknown constants and to simplify the model
by writing the input-output relationship as 
\begin{equation}
P_{out}=\frac{\theta_{3}P_{in}+\theta_{2}}{P_{in}+\theta_{1}}-\frac{\theta_{2}}{\theta_{1}},\label{eq:Alouni}
\end{equation}
with $\theta_{1}$, $\theta_{2}$, and $\theta_{3}$ being constants
obtained by curve fitting. Similar to the sigmoid model, the rational
function models mainly considers the saturation behavior of the rectifier
focusing on Region 2 and Region 3, whereas the multi-tone optimization
problem becomes interesting in Region 1. 
\section{Waveform Optimization for Maximum Power Transfer using Two Non-Linear
Rectifier Models }

In this section, we formulate the optimization problem of designing
the multi-tone signal with the goal of maximizing the harvested power
from the energy harvester without accounting for the co-existing communication
link. For simplicity, we assume at first that the power transmitter
is equipped with one antenna ($M=1$) and then the extension to the
case where $M>1$ is discussed at the end of this section.

\subsection{Single-antenna Case (M=1)}

For $M=1$, we suppress the dependence on $m$, and the transmitted
power signal becomes
\begin{alignat}{1}
x_{p}(t) & =\sum_{n=1}^{N}s_{n}\cos\left(2\pi f_{n}t+\phi_{n}\right)=\mathcal{\Re}\mathit{\left\{ \sum_{n=1}^{N}s_{n}e^{j\mathrm{2}\pi f_{n}t}e^{j\phi_{n}}\right\} ,}\label{eq:}
\end{alignat}
while the received signal at the energy harvester due to the power
transmitter becomes
\begin{alignat}{1}
y_{p}(t) & =\mathcal{\Re}\mathit{\left\{ \sum_{n=1}^{N}s_{n}h_{n}e^{j\mathrm{2}\pi f_{n}t}e^{j\phi_{n}}e^{j\psi_{n}}\right\}}=\sum_{n=1}^{N}s_{n}h_{n}\cos\left(2\pi f_{n}t+\phi_{n}+\psi_{n}\right).\label{eq:M1RecPow}
\end{alignat}

According to \protect\cite{POWValenta,WFDesign,MaxDCEH}, the performance
of the energy harvester is maximized when all tones have a zero phase
when received at the rectifier. Assuming the CSI is known at the power
transmitter, the optimum phase for the transmitted waveform will be
$\phi_{n}^{*}=-\psi_{n}$. Therefore, the received signal at the energy
harvester is
\begin{equation}
y_{p}(t)=\mathcal{\Re}\mathit{\left\{ \sum_{n=1}^{N}s_{n}h_{n}e^{j\mathrm{2}\pi f_{n}t}\right\} =}\sum_{n=1}^{N}s_{n}h_{n}\cos\left(2\pi f_{n}t\right).\label{eq:M1Yp}
\end{equation}

From (\protect\ref{eq:M1Yp}) we can write $\mathbb{E}\left\{ y_{p}(t)^{2}\right\} $
and $\mathbb{E}\left\{ y_{p}(t)^{4}\right\} $ which are found in
(\protect\ref{eq:fdc}) as 
\begin{equation}
\mathbb{E}\left\{ y_{p}(t)^{2}\right\} =\frac{1}{2}\sum_{i=1}^{N}s_{i}^{2}h_{i}^{2},\label{eq:second harmonic}
\end{equation}
\begin{equation}
\mathbb{E}\left\{ y_{p}(t)^{4}\right\} =\frac{3}{8}\sum_{i=1}^{N}s_{i}^{4}h_{i}^{4}+\frac{3}{4}\sum_{i=1}^{N}\sum_{j=1,\,j\neq i}^{N}s_{i}^{2}h_{i}^{2}s_{j}^{2}h_{j}^{2}.\label{eq:Fourth order}
\end{equation}

\subsubsection{Waveform optimization using diode non-linear model}

Following the non-linear diode model discussed in Section II, the
amplitudes of the multi-tone signal can be obtained as solutions to
the following optimization problem: 
\begin{equation}
\begin{array}{cc}
\underset{\mathbf{s}}{\max\:} & f_{DC}(\mathbf{s},\boldsymbol{\phi^{*})}\\
\textrm{subject\:to}\: & \frac{1}{2}\left\Vert \mathbf{s}\right\Vert ^{2}\leq P\thinspace.
\end{array}\label{eq:max1}
\end{equation}
Here $f_{DC}(\mathbf{s},\boldsymbol{\phi^{*}})$ is obtained from
(\protect\ref{eq:fdc}) by plugging in (\protect\ref{eq:second harmonic}) and (\protect\ref{eq:Fourth order}).
The resulting expression for $f_{DC}(\mathbf{s},\boldsymbol{\phi^{*}})$
becomes
\begin{equation}
\begin{aligned}f_{DC}(\mathbf{s},\boldsymbol{\phi^{*}})=  k_{4}R_{ant}^{2}\left(\frac{3}{8}\sum_{i=1}^{N}s_{i}^{4}h_{i}^{4}+\frac{3}{4}\sum_{i=1}^{N}\sum_{j=1,\,j\neq i}^{N}s_{i}^{2}h_{i}^{2}s_{j}^{2}h_{j}^{2}\right)
  +k_{2}R_{ant}\left(\frac{1}{2}\sum_{i=1}^{N}s_{i}^{2}h_{i}^{2}\right).
\end{aligned}
\label{eq:FdcFullexpression}
\end{equation}
Define $\mathbf{Q}=[Q_{ij}]$ such that $Q_{ij}=2\left(2-\delta_{ij}\right)\frac{3k_{4}R_{ant}^{2}}{8}h_{i}^{2}h_{j}^{2}$,
$\mathbf{a=1}$ and $b=2P$. We also let $\mathbf{x}=\left[s_{1}^{2},\,s_{2}^{2},..,\,s_{N}^{2}\right]^{T}$,
and $\mathbf{f}=\frac{k_{2}R_{ant}}{2}\left[h_{1}^{2},\,h_{2}^{2},..,\,h_{N}^{2}\right]^{T}.$
The problem can be re-written in the form of a non-homogeneous linearly
constrained quadratic program (LCQP) as follows:
\begin{equation}
\begin{array}{ccc}
\underset{\mathbf{x}}{\max}\: & \mathbf{\frac{\mathrm{\mathrm{1}}}{\mathrm{2}}x^{\mathit{T}}Qx+f^{\mathit{T}}x}\\
\textrm{ subject\:to}\: & \mathbf{a^{\mathit{T}}x}\leq b,\:\mathbf{x}\geq\boldsymbol{0}.
\end{array}\label{eq:QPGen}
\end{equation}
 Since $\mathbf{Q}$ is a nonnegative matrix and not necessarily positive
semidefinite, the problem is a non-convex LCQP which is known to be
$\mathcal{NP}$ hard \protect\cite{Burer2006}. However, for the special case
of having a convex quadratic objective function with linear constraints
and bounded inputs, this problem can be solved globally using branch
and bound methods \protect\cite{Burer2006,Chen2012} or integer linear programming
\protect\cite{2015arXivOPT}. We propose two solutions to the above optimization
problem using both methods and show by simulations that the obtained
solutions are the same within numerical precision of the simulations;
see Section VI.

\emph{Solution using integer linear programming techniques:}

The algorithm proposed in \protect\cite{2015arXivOPT} is used as a basis
for our proposed solution for (\protect\ref{eq:QPGen}). In the proposed method,
the quadratic program (QP) is reformulated as a linear complementary
problem by applying the KKT conditions. Then the complementary constraints
are reformulated using binary variables transforming the optimization
problem to a mixed-integer linear problem (MILP) which can be solved
effectively using existing MILP solvers such as CPLEX to obtain the
QP globally optimal solution.
 
For the QP in (\protect\ref{eq:QPGen}), the Lagrange multipliers $\mu\in\mathbb{R}$
are introduced for the inequality constraint and $\mathbf{\boldsymbol{\lambda}}\in\mathbb{\mathbb{R^{\mathrm{\mathit{N}}}}}$
for the non-negativity constraints. The KKT conditions for the problem
can be written as 
\begin{align}
\mathbf{Qx}+\mathbf{f}+\mathbf{a}\text{\ensuremath{\mu}}-\boldsymbol{\lambda}=0\label{eq:1a}\\
\mathbf{x}^{T}\boldsymbol{\lambda}=0\label{eq:1b}\\
\mu\left(\mathbf{\mathbf{a^{\mathit{T}}x}}-b\right)=0\label{eq:1c}\\
\mathbf{x}\geq\boldsymbol{0},\,\boldsymbol{\lambda\geq0},\thinspace\mu\thinspace\geq0.\label{eq:1d}
\end{align}
Since the problem is non-convex, these are only first order necessary
conditions for the optimal solution of (\protect\ref{eq:QPGen}). The KKT conditions
can be added as constraints the original problem in (\protect\ref{eq:QPGen})
to obtain the following equivalent\footnote{Two optimization problems are equivalent if the solution to one can
be obtained from the other, and vice-versa.} formulation:
\begin{equation}
\begin{aligned}\max_{\mathbf{x},\,\boldsymbol{\lambda},\,\mu}\: & \mathbf{\frac{\mathrm{\mathrm{1}}}{\mathrm{2}}x^{\mathit{T}}Qx+f^{\mathit{T}}x}\\
\textrm{ subject\:to}\: & \mathbf{Qx}+\mathbf{f}+\mathbf{a}\text{\ensuremath{\mu}}-\boldsymbol{\lambda}=0\\
 & \mathbf{x}^{T}\boldsymbol{\lambda}=0,\,\mu\left(\mathbf{a^{\mathit{T}}x}-b\right)=0\\
 & \mathbf{x}\geq\boldsymbol{0},\,\boldsymbol{\lambda\geq0},\thinspace\mu\geq0.
\end{aligned}
\label{eq:QKKT}
\end{equation}
As shown in \protect\cite{Burer2006,2015arXivOPT}, the KKT conditions (\protect\ref{eq:1a})-(\protect\ref{eq:1c})
can be used to linearize the objective of (\protect\ref{eq:QKKT}) as follows,
\begin{equation}
\begin{aligned}\frac{\mathrm{1}}{2}\mathbf{x^{\mathit{T}}Q}\mathbf{x}+\mathbf{f}^{T}\mathbf{x} & =\frac{\mathrm{1}}{\mathrm{2}}\left(\mathbf{f}^{T}\mathbf{x}-\mathbf{x}^{T}\mathbf{a}\mu+\mathbf{x}^{T}\mathbf{\mathbf{\boldsymbol{\lambda}}}\right)\\
 & =\mathrm{\frac{1}{2}}\left(\mathbf{f^{\mathit{T}}x}-b\mu\right).
\end{aligned}
\label{eq:Linearization}
\end{equation}
The optimization problem becomes, 

\begin{equation}
\begin{aligned}\max_{\mathbf{x},\,\boldsymbol{\lambda},\,\mu}\: & \mathrm{\frac{1}{2}}\left(\mathbf{f^{\mathit{T}}x}-b\mu\right)\\
\textrm{ subject\:to}\: & \mathbf{Qx}+\mathbf{f}+\mathbf{a}\text{\ensuremath{\mu}}-\boldsymbol{\lambda}=0\\
 & \mathbf{x}^{T}\boldsymbol{\lambda}=0,\:\mu\left(\mathbf{\mathbf{a^{\mathit{T}}x}}-b\right)=0\\
 & \mathbf{x}\geq\boldsymbol{0},\,\boldsymbol{\lambda\geq0},\thinspace\mu\geq0.
\end{aligned}
\label{eq:ModifOpt}
\end{equation}
The difficulty of solving this problem is due to the coupling among
the variables of $\mathbf{x}^{T}\boldsymbol{\lambda}=0$. A solution
method to this problem was proposed in \protect\cite{2015arXivOPT} by solving
the following equivalent MILP:
\begin{equation}
\begin{aligned}\max_{\mathbf{x},\,\boldsymbol{\lambda},\,\mu}\: & \mathrm{\frac{1}{2}}\left(\mathbf{f^{\mathit{T}}x}-b\mu\right)\\
\textrm{ subject\:to}\: & \mathbf{Qx}+\mathbf{f}+\mathbf{a}\text{\ensuremath{\mu}}-\boldsymbol{\lambda}=0\\
 & 0\leq\mathbf{x}\leq z\mathbf{u,}\:0\leq\mathbf{\boldsymbol{\lambda}}\leq\left(1-z\right)\mathbf{v},\\
 & \mu\left(\mathbf{\mathbf{a^{\mathit{T}}x}}-b\right)=0,\:z\in\left\{ 0,1\right\} ,
\end{aligned}
\label{eq:ModifMILP}
\end{equation}
where $\mathbf{u},\mathbf{\,v}$$\in\mathbb{\mathbb{R^{\mathrm{\mathit{N}}}}}$
are the upper bounds on the the decision variables $\mathbf{x}$,
\textbf{$\mathbf{\boldsymbol{\lambda}}$} such that there are globally
optimal KKT points $\left(\mathbf{x},\,\mu,\,\mathbf{\mathbf{\boldsymbol{\lambda}}}\right)$
for the primal QP satisfying $\mathbf{x}\leq\mathbf{u}$, $\mathbf{\boldsymbol{\lambda}}\leq\mathbf{v}$.
The upper bound $\mathbf{u}$ is given by $\mathbf{u}=\mathbf{x^{*}}$,
where $\mathbf{\mathbf{x^{*}}}$ is the solution to the following
linear programming (LP) problem:
\begin{equation}
\begin{array}{ccc}
\underset{\mathbf{x}}{\max\:} & \mathbf{x}\\
\textrm{ subject\:to}\: & \mathbf{a^{\mathit{T}}x}\leq b,\:\mathbf{x}\geq\boldsymbol{0},
\end{array}\label{eq:Ubound}
\end{equation}
and $\mathbf{v}=\mathbf{\boldsymbol{\lambda^{*}}}$, with $\mathbf{\boldsymbol{\lambda^{*}}}$
being the solution to the LP problem given by
\begin{equation}
\begin{array}{ccc}
\underset{\mathbf{\boldsymbol{\lambda}}}{\max\:} & \mathbf{\boldsymbol{\lambda}}\\
\textrm{ subject\:to}\: & \mathbf{Qx}+\mathbf{f}+\mathbf{a}\text{\ensuremath{\mu}}-\boldsymbol{\lambda}+\mathbf{\mathbf{\boldsymbol{\rho}}}=0\\
 & 0\leq\mathbf{X}\leq\mathbf{u}\mathbf{u}^{T}\\
 & 0\leq\mathbf{x}\leq\mathbf{u},\thinspace\mathbf{\boldsymbol{\lambda}},\thinspace\mathbf{\mathbf{\boldsymbol{\rho}}}\geq0,\thinspace\mathbf{X\thinspace}\in\thinspace S^{N},
\end{array}\label{eq:Vbound}
\end{equation}
where $S^{N}$ represents the set of $N\times N$ real symmetric matrices.
The matrix $\mathbf{X}$ represents the linearization of the matrix
$\mathbf{x}\mathbf{x}^{T}\in\thinspace S^{N}$, and $\mathbf{\boldsymbol{\rho}}\in\mathbb{\mathbb{R^{\mathrm{\mathit{N}}}}}$
is a dual variable introduced to calculate the bound. Solving the
LP in (\protect\ref{eq:Ubound}) and (\protect\ref{eq:Vbound}), the obtained bounds
are then applied to the MILP in (\protect\ref{eq:ModifMILP}), and then solving
this equivalent problem using MILP solvers such as CPLEX \protect\cite{2015arXivOPT} yields to
obtaining an optimal solution to the primal problem given by (\protect\ref{eq:QPGen}). 

\emph{Solution using branch-and bound-technique:}

A globally optimal solution to non-convex quadratic programs using
a finite branch-and-bound method was presented in \protect\cite{Burer2006,Chen2012}.
Optimality is achieved by enforcing the first-order KKT conditions
through branching and then solving a convex relaxation of the quadratic
program at each node of the branch-and-bound tree. Two types of relaxations
were discussed by Burer and Vanderbussche \protect\cite{Burer2006}, namely
Linear program (LP) relaxation and a semidefinite program (SDP) relaxation.
We will start by describing the solution using the LP relaxation approach
and then describe the SDP formulation. 

The feasible set of (\protect\ref{eq:QPGen}) is referred to as 
\begin{equation}
{\cal P}:=\left\{ \mathbf{x}\in\mathbb{\mathbb{R^{\mathrm{\mathit{N}}}}}:\mathbf{a^{\mathit{T}}x}\leq b,\,\mathbf{x}\geq\mathbf{0}\right\} .\label{eq:feasibleset}
\end{equation}
The solution starts by introducing the the Lagrange multipliers $\mu\in\mathbb{R}$
for the inequality constraint and $\mathbf{\boldsymbol{\lambda}}\in\mathbb{\mathbb{R^{\mathrm{\mathit{N}}}}}$
for the non-negativity constraints of the QP in (\protect\ref{eq:QPGen})
and then writing the first-order KKT conditions in the same way as
(\protect\ref{eq:1a})-(\protect\ref{eq:1d}). Then, two sets ${\cal G}$ and ${\cal {\cal C}}$
are defined as:
\begin{equation}
{\cal G}:=\left\{ \mathbf{\left(\mu,\mathbf{\boldsymbol{\lambda}}\right)}\geq0\,:\boldsymbol{\lambda}-\mathbf{a}\text{\ensuremath{\mu}}=\mathbf{Qx}+\mathbf{f}\right\} ,\label{eq:SetG}
\end{equation}
\begin{equation}
{\cal C}:=\left\{ \mathbf{\left(\mu,\mathbf{\boldsymbol{\lambda}}\right)}\geq0\,:\mu\left(\mathbf{a^{\mathrm{\mathit{T}}}x}-b\right)=0,\,\mathbf{x}^{T}\boldsymbol{\lambda}=0\right\} .\label{eq:SetC}
\end{equation}
Any local optimal solution $\mathbf{x}$ must  satisfy ${\cal G}\cap{\cal C}\neq\emptyset.$
Following (\protect\ref{eq:Linearization}), the optimization problem can
be reformulated in a similar way to (\protect\ref{eq:ModifOpt}) as 
\begin{equation}
\begin{aligned}\underset{\mathbf{x}}{\max\:} & \mathrm{\mathrm{\frac{1}{2}}}\left(\mathbf{f^{\mathit{T}}x}-b\mu\right)\\
\textrm{ subject\thinspace to}\thinspace\thinspace & \mathbf{x}\in{\cal P},\thinspace\mathbf{\left(\mu,\mathbf{\boldsymbol{\lambda}}\right)}\in{\cal G}\cap{\cal C.}
\end{aligned}
\label{eq:BBKKT}
\end{equation}
The formulation in (\protect\ref{eq:BBKKT}) allows the use of finite branch-and-bound,
where the complementary constraints are enforced using linear equations.
At each node of the tree, two index sets are introduced ${\cal F}^{x}$,
${\cal F}^{\lambda}\subseteq\left\{ 1,...,n\right\} $ which satisfy
${\cal F}^{x}\cap{\cal F}^{\lambda}=\emptyset.$ These two index sets
are used to enforce the complementary constraints by solving the following
LP optimization problem:

\begin{equation}
\begin{aligned}\underset{\mathbf{x}}{\max\:} & \mathrm{\mathrm{\frac{1}{2}}}\left(\mathbf{f^{\mathit{T}}x}-b\mu\right)\\
\textrm{ subject\thinspace to}\thinspace\thinspace & \mathbf{x}\in{\cal P},\thinspace\mathbf{\left(\mu,\mathbf{\boldsymbol{\lambda}}\right)}\in{\cal G}\cap{\cal C}\\
 & x_{i}=0,\,i\in{\cal F}^{x}\\
 & \lambda_{i}=0,\,i\in{\cal F}^{\lambda}\\
 & \mathbf{a^{\mathrm{\mathit{T}}}x}-b=0,\thinspace \mu=0.
\end{aligned}
\label{eq:BBlp}
\end{equation}
The constraint $\mathbf{\left(\mu,\mathbf{\boldsymbol{\lambda}}\right)}\in C$
can be dropped to obtain an LP relaxation. This relaxation maintains
complementary because $x_{i}\lambda_{i}=0$ for all $i\in{\cal F}^{x}\cup{\cal F}^{\lambda}$.
Branching on a node is done by selecting some $i\in\left\{ 1,...,n\right\} \setminus{\cal F}^{x}\cup{\cal F}^{\lambda}$
and then creating two children by adding $i$ to ${\cal F}^{x}$ for
one child and $i$ to ${\cal F}^{\lambda}$ for the other \protect\cite{Burer2006,Burer2009}. Adding
indices progressively enforces more complementarity at nodes deeper
in the tree. It should be noted that the root node in the branch tree
has all index sets empty. 

SDP relaxations are known to be bounded. That is why they are preferred
to LP relaxations in finite branch-and-bound algorithms. Instead of
solving a LP at each node, the QP in (\protect\ref{eq:QPGen}) is reformulated
in the form of a relaxed SDP as follows, 

\begin{equation}
\begin{aligned}\max_{\mathbf{X},\thinspace\mathbf{x}}\,\thinspace & \frac{1}{2}\textrm{tr}\left(\mathbf{QX}\right)\mathbf{+f^{\mathit{T}}x}\\
\textrm{ subject\thinspace to}\thinspace\thinspace & \mathbf{a^{\mathit{T}}x}\leq b,\:\mathbf{X}\succeq\mathbf{x}\mathbf{x^{\mathit{T}},\:\mathbf{X}\geq\mathbf{0},\:\mathbf{x}\geq\mathbf{0},}
\end{aligned}
\label{eq:QSDP}
\end{equation}
where $\mathbf{X}\geq\mathbf{0}$ denotes a positive semidefinite
matrix. The complementary constraints are then added to the above
SDP as in (\protect\ref{eq:BBlp}) and enforced through branching. Finite
branch-and-bound using SDP was proven to globally solve non-convex
QCQP \protect\cite{Burer2006}. However, it has high computational complexity
due to solving a SDP at each node of the BB tree. On the other hand,
in the MILP technique, the dual bounds are calculated and the equivalent
problem is solved using CPLEX once. Since the work proposed in \protect\cite{2015arXivOPT}
is still under review, we implemented both solution methods to show
that the MILP technique can be used to obtain a globally optimal solution
for our multi-tone optimization problem. 

\textbf{Computational Complexity:} Evidently, obtaining globally optimal
solutions for a non-convex problem comes at the cost of elevated complexity
for the algorithms. Here, we make approximate complexity calculations
for both the BB and MILP solutions. The computational complexity of
the BB methods is dominated by solving the SDP, in (\protect\ref{eq:QSDP}):
this can be done using generic solvers with $\approx\mathcal{O}(N^{4})$,
for each node in the BB tree. Assuming a total maximum of BB nodes
of $K$, the total complexity of the BB algorithm is $\approx K\mathcal{O}(N^{4})$
operations. On the other hand, obtaining similar complexity calculations
is more challenging for our MILP solution, since the complexity of
MILP is harder to characterize in closed-form expressions. We note
that complexity reduction techniques are interesting future work,
namely, interior point method to reduce the complexity of (\protect\ref{eq:QSDP}),
and heuristics to obtain locally optimal solutions to the non-convex
QP. 

\subsubsection{Waveform optimization using curve-fitting model}

The second-order polynomial model given by (\protect\ref{eq:SecondOrderP})
can be slightly modified in the case of multi-tone excitation in the
following way:
\begin{equation}
P_{out}\left(t\right)=\beta_{1}P_{in}^{2}\left(t\right)+\beta_{2}P_{in}\left(t\right)+\beta_{3},\label{eq:ModifCfit1}
\end{equation}
where $P_{in}\left(t\right)=y_{p}^{2}\left(t\right)$ is the instantaneous
input power to the energy harvester. It is required to maximize the
average output power of the energy harvester given by $P_{out}=\mathbb{E}\left\{ P_{out}\left(t\right)\right\} $.
By averaging Eq. (\protect\ref{eq:ModifCfit1}) the output DC power of the
rectifier can be written as:
\begin{equation}
P_{out}=\beta_{1}\mathbb{E}\left\{ y_{p}(t)^{4}\right\} +\beta_{2}\mathbb{E}\left\{ y_{p}(t)^{2}\right\} +\beta_{3},\label{eq:OptimizPloyfit}
\end{equation}
with $\mathbb{E}\left\{ y_{p}(t)^{2}\right\} $, $\mathbb{E}\left\{ y_{p}(t)^{4}\right\} $
expressed in the same way as (\protect\ref{eq:second harmonic}) and (\protect\ref{eq:Fourth order}),
respectively. Following the assumption of perfect CSI at the power
transmitter and substituting by the optimal phases given by $\phi_{n}^{*}=-\psi_{n}$
, the multi-tone optimization problem in the case of the modified
second order polynomial model based becomes:

\begin{align}
\max_{\mathbf{s}}\thinspace & \beta_{1}\left(\frac{3}{8}\sum_{i=1}^{N}s_{i}^{4}h_{i}^{4}+\frac{3}{4}\sum_{i=1}^{N}\sum_{j=1,\,j\neq i}^{N}s_{i}^{2}h_{i}^{2}s_{j}^{2}h_{j}^{2}\right)\nonumber
 +\beta_{2}\left(\frac{1}{2}\sum_{i=1}^{N}s_{i}^{2}h_{i}^{2}\right)\label{eq:OptQfit}\\
\textrm{subject\thinspace to}\thinspace\, & \frac{1}{2}\left\Vert \boldsymbol{\mathbf{s}}\right\Vert ^{2}\leq P.\nonumber 
\end{align}
Again, defining \textbf{$\mathbf{x}=\left[s_{1}^{2},\,s_{2}^{2},..,\,s_{N}^{2}\right]^{T}$},
the problem can be written in the form of a non-convex LCQP in a similar
way to (\protect\ref{eq:QPGen}) and solved as described before.

\subsection{Extension to MISO Power Transfer ($M>1$)}

When $M>1$, the expressions given in (\protect\ref{eq:second harmonic})
and (\protect\ref{eq:Fourth order}) do not hold and the problem in this case
cannot be formulated as a LCQP. However, a simple trick can be applied
to decouple the frequency and the space domains without affecting
the performance \protect\cite{WFDesign}. Noticing that the optimal phases
for maximizing the output of the energy harvester are given by $\phi_{n,m}^{*}=-\psi_{n,m}$,
one can generalize this to express the optimal weight vector represented
by $\mathbf{w}_{n}$ as the matched beamformer given by 
\begin{equation}
\mathbf{w}_{n}=s_{n}\frac{\mathbf{h}_{n}^{H}}{\left\Vert \mathbf{h}_{n}\right\Vert }.\label{eq:MBF}
\end{equation}
Applying the weight vector from (\protect\ref{eq:MBF}) to (\protect\ref{eq:RecPowerEh}),
the MISO channel is converted to the equivalent scalar channel and
the received signal at the energy harvester becomes
\begin{equation}
y_{p}(t)=\mathcal{\Re}\mathit{\left\{ \sum_{n=1}^{N}s_{n}\left\Vert \mathbf{h}_{n}\right\Vert e^{j\omega_{n}t}\right\} }.\label{eq:DecopSpandFreq}
\end{equation}
The transmitter power constraint remains the same as $\frac{1}{2}\left\Vert \mathbf{s}\right\Vert ^{2}\leq P$.
This allows the formulation of the optimization problem as a LCQP
by replacing each $h_{n}$ in (\protect\ref{eq:second harmonic}) and (\protect\ref{eq:Fourth order})
by $\left\Vert \mathbf{h}_{n}\right\Vert $ which can be interpreted
as the effective channel gain at frequency $n$. The optimal amplitude
of the $n^{th}$ tone given by $s_{n}$ is then obtained by solving
the LCQP as described previous subsections either by applying the
diode model or the curve-fitting model.

\section{Waveform Optimization for In-Band SWIPT}

\subsection{Problem Motivation}

\begin{figure}[tb!]
\begin{centering}
\includegraphics[width=3in,height=1.5in]{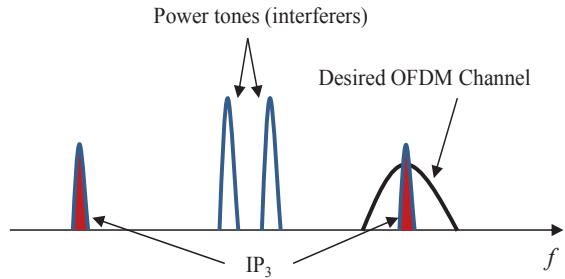}
\par\end{centering}
\caption{Distortion and blocking due to third-order inter-modulation products.
\label{fig:Distortion-and-blocking} }
\vspace{-0.2in}
\end{figure}

In this section, we consider the co-existing communication link described
in Fig. \protect\ref{fig:BlockDigaram}. One of the important building blocks
in the RF front end of any information receiver is the LNA which amplifies the received RF signal before the mixing and
filtering stages to extract the baseband signal. LNAs are active components
having a non-linear behavior. However, they are designed to operate
linearly within a certain power range corresponding to the received
communication signals. At power levels higher than the specific design
range, the LNA of the information receiver saturates and exhibits
a non-linear behavior producing inter-modulation products and high-order
harmonics. For example, if two signals are received at frequencies
$f_{1}$ and $f_{2}$ , the inter-modulation products appear at frequencies
$f_{i,j}=\pm if_{1}\pm jf_{2}$ where $i,\thinspace j$ are positive
integers. The order of the inter-modulation product is defined by
$i+j$. Usually, the third order inter-modulation products given by
$2f_{1}-f_{2}$ and $2f_{2}-f_{1}$ are the most problematic \protect\cite{pozar2011microwave}.
This is because if $f_{1}$ and $f_{2}$ are close to each other,
the third order inter-modulation products may fall inside the channel
of interest with a high power level that is enough to block it as
shown in Fig. \protect\ref{fig:Distortion-and-blocking}. As power signals
are usually transmitted at a higher power level than communication
signals, the sub-carrier frequencies in the multi-tone signal are
chosen such that they do not interfere with any information sub-carrier
or interfere at most with one sub-carrier as discussed in \protect\cite{WFOFDM}.
However, from the experiment described in the introduction, this was
found to be not enough to preserve the communication link and the
data rates. In case of in-band SWIPT, the inter-modulation products
of the multi-tone signal are more likely to fall inside the channel
of interest causing interference. Therefore, it is necessary
to prevent the LNA of the information receiver from saturation and
keep it in its linear operation region.

In RF applications, the input impedance of active components changes
significantly with the input power, which makes the matching effect
more critical. For this reason, it is more common to express the constraints
on the non-linearity of RF amplifiers in terms of power constraints
instead of voltage constraints, namely the 1-dB compression point
($P_{1\mathrm{dB}}$) and the third-order intercept ($IP_{3}$) point.
According to \protect\cite{pozar2011microwave}, $P_{1\mathrm{dB}}$ is defined
as the input power at which the amplifier gain drops by 1 dB from
the normal linear gain. It provides a value for the input power at
which the amplifier goes into compression and becomes saturated. The
$IP_{3}$ point indicates the value of the input power at which the
amplitude of the third-order inter-modulation products becomes equal
to the input signal. For a LNA to maintain its linear behavior, the
input power should be below the 1-dB compression point (a typical
value for $P_{1\mathrm{dB}}$ at 2.4 GHz is around -15 dBm) and sufficiently
lower than the $IP_{3}$ point. 

\subsection{Formulating the Optimization Problem}

\subsubsection{Single antenna case }

Based on the above discussion, we add a constraint on the received
power at the nearby communication receiver $P_{sat}$ to prevent its LNA from saturation. Considering the case where $M=1$,
the received signal at the information receiver due to the power transmitter
from (\protect\ref{eq:Yirec}) becomes
\begin{equation}
y_{IP}(t)=\mathcal{\Re}\mathit{\left\{ \sum_{n=1}^{N}s_{n}g_{n}e^{j\mathrm{2}\pi f_{n}t}e^{j\phi_{n}}e^{j\zeta_{n}}\right\} }.\label{eq:YIrecM1}
\end{equation}
Therefore, the average power received is given by $\mathbb{E}\left\{ y_{IP}(t)^{2}\right\} $
and can be written as $\frac{1}{2}\sum_{n=1}^{N}s_{n}^{2}g_{n}^{2}$.
Adding the received power constraint\footnote{It should be noted that the received power due to the legitimate information
	link also contributes towards the LNA saturation. Therefore, the constraint
	$P_{sat}$ should actually be lower than $P_{1\mathrm{dB}}$, i.e.,
	$P_{sat}=P_{1\mathrm{dB}}-\varepsilon$, where $\varepsilon$ represents
	the average power received due to the information signal.} to the optimization problem results
in,
\begin{equation}
\begin{array}{cc}
\underset{\boldsymbol{\mathbf{s}}}{\max}\: & f_{DC}(\boldsymbol{\mathbf{s}},\boldsymbol{\mathbf{\phi}^{*})}\\
\textrm{ subject\:to}\: & \frac{1}{2}\left\Vert \boldsymbol{\mathbf{s}}\right\Vert ^{2}\leq P,\:\frac{1}{2}\sum_{n=1}^{N}s_{n}^{2}g_{n}^{2}\leq P_{sat}.
\end{array}\label{eq:Max2}
\end{equation}
After reformulating the problem in the form of LCQP we re-write the
optimization problem as
\begin{equation}
\begin{aligned}\max_{\mathbf{x}}\: & \frac{1}{2}\mathbf{xQx^{T}+f^{T}x}\\
\textrm{ subject\:to}\: & \left[\begin{array}{c}
\mathbf{1^{T}}\\
\mathbf{v}
\end{array}\right]\mathbf{x}\leq\left[\begin{array}{c}
2P\\
2P_{sat}
\end{array}\right]\\
 & \mathbf{\:x}\geq\mathbf{0},
\end{aligned}
\label{eq:InfoConstQP2}
\end{equation}
where $\mathbf{v}=\left[g_{1}^{2},g_{2}^{2},..,g_{n}^{2}\right]$.
By defining $\mathbf{\boldsymbol{A_{2}}=\left[\begin{array}{c}
\mathbf{1^{T}}\\
\mathbf{v}
\end{array}\right]}$ and $\mathbf{b_{2}}=\left[\begin{array}{cc}
2P & 2P_{sat}\end{array}\right]^{T}$, the problem can be solved in the same procedure described in Section
III.

\subsubsection{MISO case ($M>1$)}

Although the matched beamformer given by (\protect\ref{eq:MBF}) maximizes
the received power at the energy harvester, it does not take into
account co-existing information transmitters and receivers. The received
signal at the nearby information receiver after applying this matched
beamformer becomes
\begin{equation}
y_{IP}(t)=\mathcal{\Re}\mathit{\left\{ \sum_{n=1}^{N}s_{n}\mathbf{g}_{n}\frac{\mathbf{h}_{n}^{H}}{\left\Vert \mathbf{h}_{n}\right\Vert }e^{j2\pi f_{n}t}\right\} }.\label{eq:MISOSWIPT}
\end{equation}
As the number of transmitting antennas $M$ increases, the radiated
beam for power transmission becomes narrower and more directed towards
the energy harvester, thus, limiting the amount of power received
at the information receiver. This can also be viewed from the value
of $\mathbb{E}\left\{ \mathbf{g}_{n}\mathbf{h}_{n}^{H}\mathbf{h}_{n}\mathbf{g}_{n}^{H}\right\} $
which tends to zero as $M$ increases (by the law of large numbers).
Nevertheless, the possible interference resulting from the matched
beamformer is still captured in the constraint given by $\mathbb{E}\left\{ y_{IP}(t)^{2}\right\} \leq P_{sat}.$
One may also think of a hybrid beamforming scheme in which the weighting
vector is a linear combination of the matched beamformer and the zero-forcing
beamformer as in MISO interference channels \protect\cite{ErikLarssonMISO,IoannisMISO}.
However, by expressing the weighting vector as a linear combination
of these two beamformers, the multi-tone optimization problem becomes
intractable and cannot be formulated as a LCQP. The problem of global
optimal multi-tone multi-antenna SWIPT remains interesting for future
research. 

\section{Simulation Results}
Consider a WPT system with $N=8$ tones. We start by assuming a SISO
system with $M=1$ and comparing our MILP and BB solutions to the
multi-tone optimization problem to three strategies for power allocation
to different tones. The first strategy is the equal amplitude splitting
(Equal AS), i.e., to divide power equally among all tones such that
$s_{n}=\sqrt{\frac{2P}{N}}$. The second strategy is the maximum-ratio
transmission (MRT) in which the amplitude of the $n^{th}$ tone is
given by $s_{n}=h_{n}\sqrt{\frac{2P}{\sum_{i=1}^{N}h_{i}^{2}}}$.
The third is the adaptive single sinewave, in which all the power
is allocated to the tone corresponding to the strongest channel $h_{n^{*}}$
with $n^{*}=\arg\max_{n}h_{n}$, i.e., $s_{n^{*}}=\sqrt{2P}$. We
also compare our solution to the one proposed in \protect\cite{WFDesign}
using reverse geometric programming (Reverse GP). Details of reverse
geometric programming solution method are omitted from this paper,
since it was implemented only for comparison reasons. We consider
a Rician channel model motivated by the fact that for efficient WPT
systems LOS should exist between the wireless power transmitter and
the energy harvester. We modified the channel model in \protect\cite{SWIPTErikLarss}
to be $h_{n}=\sqrt{L_{P}}Z$, where $Z\thicksim\mathcal{CN}\left(\sqrt{\frac{\kappa}{\kappa+1}},\frac{1}{\kappa+1}\right)$.
Here, $L_{P}$ represents the path loss given by $\left(\frac{\lambda}{4\pi d}\right)^{2}$
with $d$ being the distance between the transmitter and the receiver,
$\lambda$ is the wavelength at the carrier frequency, and $\kappa$
is the Rice factor. We assume that $\kappa_{\left(dB\right)}=3$ dB
and that the channel frequency responses $h_{n}$ are independent.
The antenna gains are eliminated from the considered channel model
and assumed to be unity. We assume that $f_{o}=2.4$ GHz, $\Delta f_{p}=1.25$
MHz and the energy harvester is at distance $d=8\lambda$ from the
transmitter. We assume that $R_{ant}=50\,\Omega$. We first present
the results for maximizing the output of the energy harvester using
the diode non-linear model and the curve-fitting model. Then, we consider
the co-existing communication link. 

The power levels chosen for simulations in this paper are within the
the Federal Communications Commission (FCC) safety regulations. For
example, FCC allows a maximum of $4\,\mathrm{W}$ equivalent isotropic
radiated power (EIRP) on the 2.4 GHz band. This permits the reception
of power that can reach higher than $100\,\mathrm{\mu W}$ at an energy
harvester placed at distance $8\lambda.$ The regulations vary in
different countries. In Sweden, for example, the maximum EIRP allowed
at 2.4 GHz is $100\,\mathrm{mW}$. However, a maximum EIRP of $500\,\mathrm{mW}$ is
allowed for powering RFID tags in the band 2.446 GHz to 2.454 GHz.
These regulations have been taken care of when conducting experiments
such as the one mentioned in the introduction to this paper. It is
worth mentioning that all the current standardization is defined for
communication networks but no standardization has been defined for
wireless power transfer so far. 

\subsection{Results for Maximizing the Output of the Energy Harvester }

\subsubsection{Diode model}

The diode characteristics are obtained from the SMS-7630 data sheet.
This diode, provided by Skyworks, is known to have a low turn-on voltage
(i.e., it can be as low as 60 mV) which minimizes the power consumption,
and hence, improves the RF-to-DC conversion efficiency at low-input
power levels. Following this, we assume $i_{s}=5\,\mathrm{\mu A}$,
$v_{t}=25.86\,\mathrm{mV}$, and $\gamma=1.05$. For implementing
the reverse geometric programming method, the relative error threshold
was selected to be $\epsilon=10^{-6}$. 

\begin{figure}[t]
\begin{centering}
\subfloat[Normalized channel coefficients.\label{fig:Normalized-channel-frequency diode}]{\includegraphics[width=3.5in,height=2.5in]{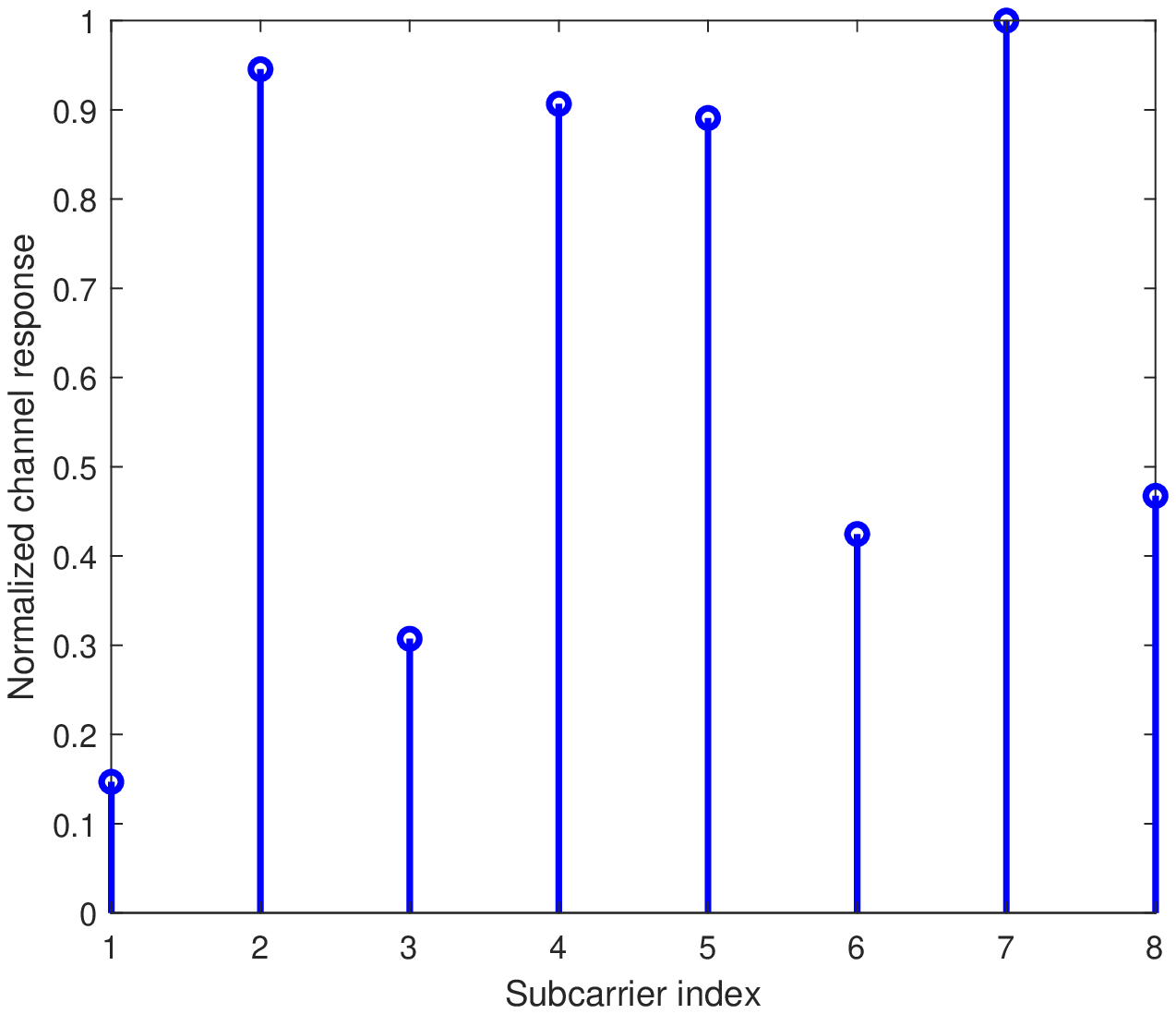}}
\subfloat[Normalized mutli-tone amplitudes.\label{fig:Mutli-tone-amplitudes}]{\includegraphics[width=3.5in,height=2.5in]{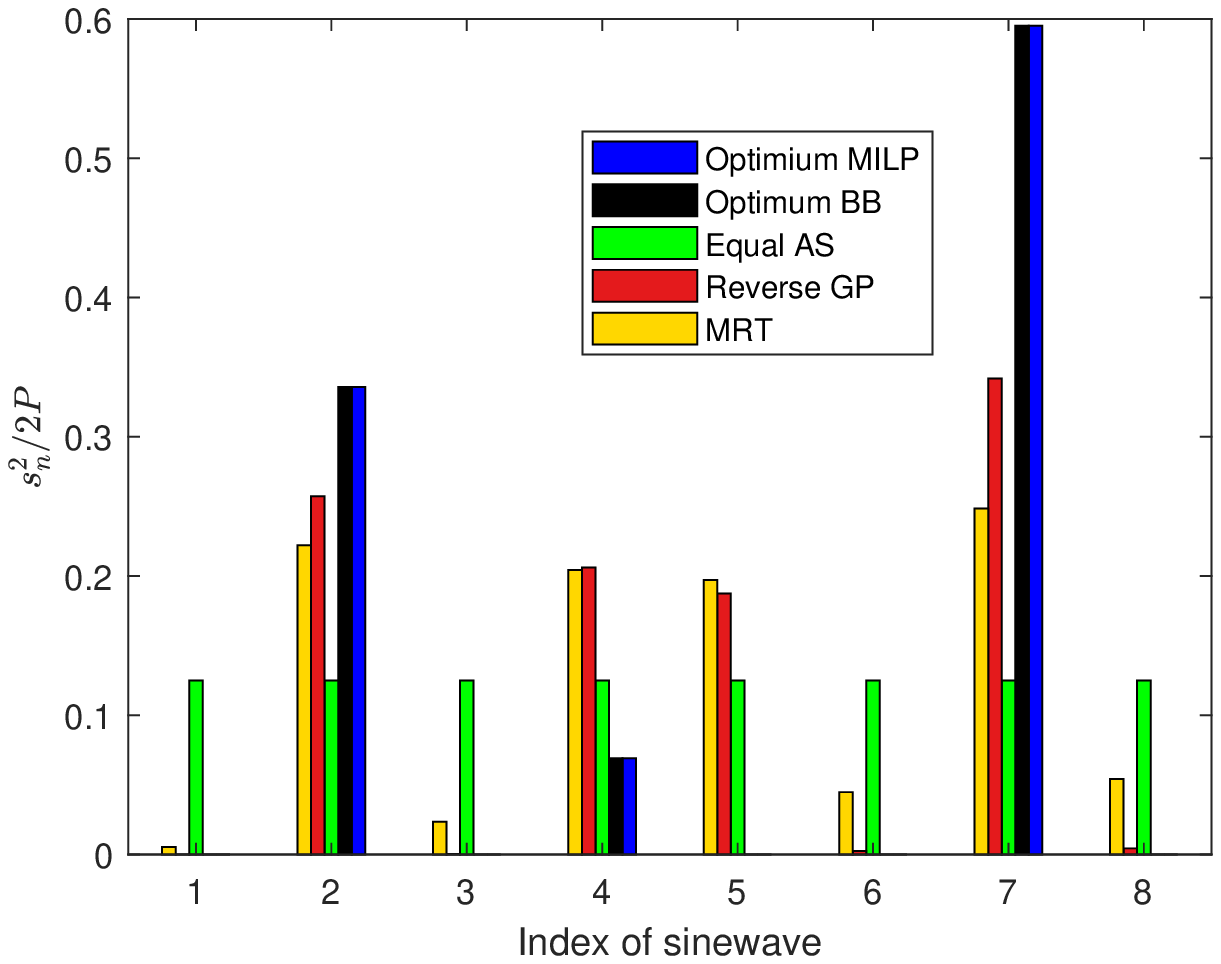}} \caption{Power allocation to different tones resulting from the diode model for one possible channel realization at $P_{EH}=50\,\mathrm{\mu W}$.}
\par\end{centering}
\vspace{-0.2in}
\end{figure}

\begin{figure}[t]
\begin{centering}
\includegraphics[width=3.5in,height=2.5in]{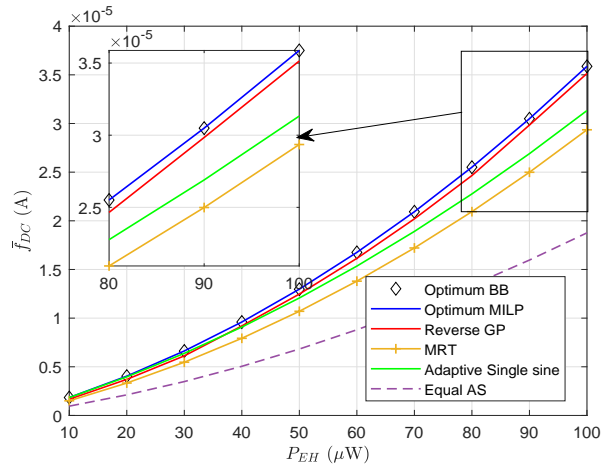}
\par\end{centering}
\caption{Average output current (over many channel realizations) represented
by $\bar{f}_{DC}$ vs. transmitted power (normalized to path loss)
for $M=1$.\label{fig:Output-Current-Diode}}
\vspace{-0.2in}
\end{figure}

In Fig. \protect\ref{fig:Normalized-channel-frequency diode} , the frequency
response (normalized to $h_{max}$) of one possible realization of
the assumed channel is plotted. We normalize the transmitted power
$P$ to the path loss $L_{P}$ to represent the maximum possible received
power at the energy harvester and denote it by $P_{EH}$. For this
particular realization shown in Fig. \protect\ref{fig:Normalized-channel-frequency diode}
and for $P_{EH}=50\,\mathrm{\mu W}$, the power allocated to each
tone resulting from the different power allocation strategies is shown
in Fig. \protect\ref{fig:Mutli-tone-amplitudes} normalized to the total power
given by $P$. It can be observed that the optimal strategy tends
to allocate power to tones having the strongest channels. Both optimal
schemes show the same results; power is divided proportionally among
the three strongest tones. It should also be mentioned that as the
transmitted power increases, the optimal schemes starts allocating
power to the next strong tone. Reverse GP tends to show a behavior
similar to MRT as can be observed from the results in Fig. \protect\ref{fig:Mutli-tone-amplitudes}
and from the discussion in \protect\cite{WFDesign,Lowcomplexity}. Since the
output current is proportional to $f_{DC}$, we refer to $f_{DC}$
as the output current in the rest of this section. In Fig. \protect\ref{fig:Output-Current-Diode},
the average output current (over many channel realizations) is plotted
vs. the transmitted power for different power allocation strategies.
It can be noticed that our solution to the optimization problem achieves
the maximum possible output current for a given input power among
all other power allocation strategies. For example, at $P_{EH}=50\,\mathrm{\mu W}$,
using the optimal schemes can improve the output current from the
energy harvester by approximately $0.5\,\mathrm{\mu A}$ ($\approx4\%$
increase) compared to the Reverse GP solution. As the transmitted
power increases (accordingly $P_{EH}$ increases), higher gains can
be obtained by applying the optimal allocation strategies. It can
also be observed from both Fig. \protect\ref{fig:Mutli-tone-amplitudes} and
Fig. \protect\ref{fig:Output-Current-Diode} that the solutions obtained from
MILP and finite BB methods coincide with each other.

\begin{figure}[t]
\centering{}\includegraphics[width=3.5in,height=2.5in]{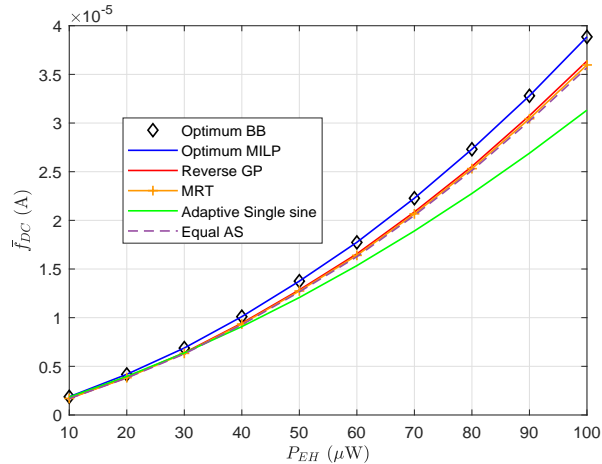}\caption{Average output current represented by $\bar{f}_{DC}$ vs. normalized
transmitted power in the case of an approximately frequency-flat Channel.\label{fig:Approximately-Frequency-Flat-Cha}}
\vspace{-0.2in}
\end{figure}

In Fig. \protect\ref{fig:Approximately-Frequency-Flat-Cha}, we evaluate the
performance of different allocation strategies with an approximately
frequency-flat channel generated by letting $h_{n}\sim\mathcal{N}\left(1,0.05\right)$.
By varying the transmitted power and observing the output DC current,
it can be noticed that MRT and Equal AS strategies have better performance
than adaptive single sine strategy, opposite to the frequency-selective
channel case. It can also be observed that the gains obtained by applying
the optimal strategies in this case are higher compared to Fig. \protect\ref{fig:Output-Current-Diode}.
The output current can be improved by approximately $1\,\mathrm{\mu A}$
($\approx7\%$ increase) at $P_{EH}=50\,\mathrm{\mu W}$ compared
to the Reverse GP solution. In the case of a perfectly flat channel
(i.e., all channel coefficients are exactly the same), the optimal
strategy, of course, is to split power equally among all tones which
is obtained from MRT, Equal AS, optimum MILP, optimum BB, and Reverse
GP strategies. 
\begin{figure}[t]
	\begin{centering}
		\includegraphics[width=3.5in,height=2.5in]{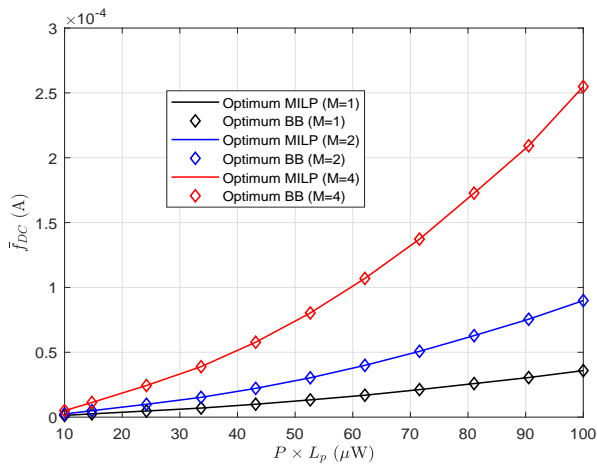}
		\par\end{centering}
	\caption{Comparison of the average output current (over many channel realizations)
		represented by $\bar{f}_{DC}$ vs. transmitted power (normalized to
		path loss) for different values of $M$. \label{fig:Output-Current-MISO}}
	\vspace{-0.2in}
\end{figure}

The effect of increasing the number of antennas at the transmitter
on the output of the energy harvester is investigated in Fig. \protect\ref{fig:Output-Current-MISO}\footnote{Due to the high complexity in implementing the reverse geometric programming
for $M>1$ and the unavailability of the code for reuse, we only show
results for MISO WPT using our optimal schemes.}. The output current of the energy harvester is plotted vs. $P\times L_{p}$
which is the same as $P_{EH}$ described for $M=1$. However, it is
more accurate to express it in this way in the MISO case since the
maximum possible received power at the energy harvester differs from
the SISO case due to the array gain (beamforming gain). As expected,
increasing the number of transmitting antennas, increases the output
current of the energy harvester. 
\subsubsection{Curve-fitting model based on circuit simulations}
\begin{figure}[t]
	\begin{centering}
		\includegraphics[width=3in,height=1.5in]{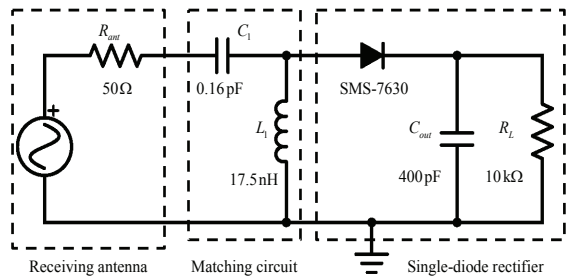}
		\par\end{centering}
	\caption{Single-diode rectifier with L-matching network.\label{fig:Single-diode-rectifier-wit}}
	\vspace{-0.2in}
\end{figure}

In order to evaluate the performance of the second-order polynomial
curve-fitting model discussed in Section IV. A single-diode rectifier
circuit was designed and simulated to achieve its maximum efficiency
at $50\,\mathrm{\mu W}$ using Keysight Technologies Advanced Design
Systems (ADS). The designed rectifier is shown in Fig. \protect\ref{fig:Single-diode-rectifier-wit}.
The choice of the output capacitor and resistor affects the low-pass
filtering characteristics of the rectifier. In order to deliver a
smooth DC output, the RC circuit time constant $\tau=R_{L}C_{out}$
should be chosen to be sufficiently larger than $\frac{1}{\Delta f_{P}}$.
The values of $C_{1}$ and $L_{1}$ are calculated according to L-matching
circuit formulas \protect\cite{pozar2011microwave} in order to match the
rectifier to the $50\,\Omega$ antenna. As mentioned before in Section
III, due to the presence of the non-linear diode, the input impedance
of the rectifier circuit varies with varying the input power to the
rectifier. It is difficult to keep a good level of matching for a
wide range of input powers with a simple design, especially at very
low-input power. That is why, the rectenna is optimized to have its
maximum efficiency at a certain input power level. Large Signal S-Parameters
(LSSP) and Harmonic Balance (HB) simulators in ADS were used to calculate
to the input impedance for matching and calculate the output DC power,
respectively.

\begin{figure}[t]
	\begin{centering}
		\subfloat[Normalized mutli-tone amplitudes obtained at $P_{EH}=50\,\mathrm{\mu W.}$\label{fig:Mutli-tone-amplitudes-Curvefit}]{
			\includegraphics[width=3.5in,height=2.5in]{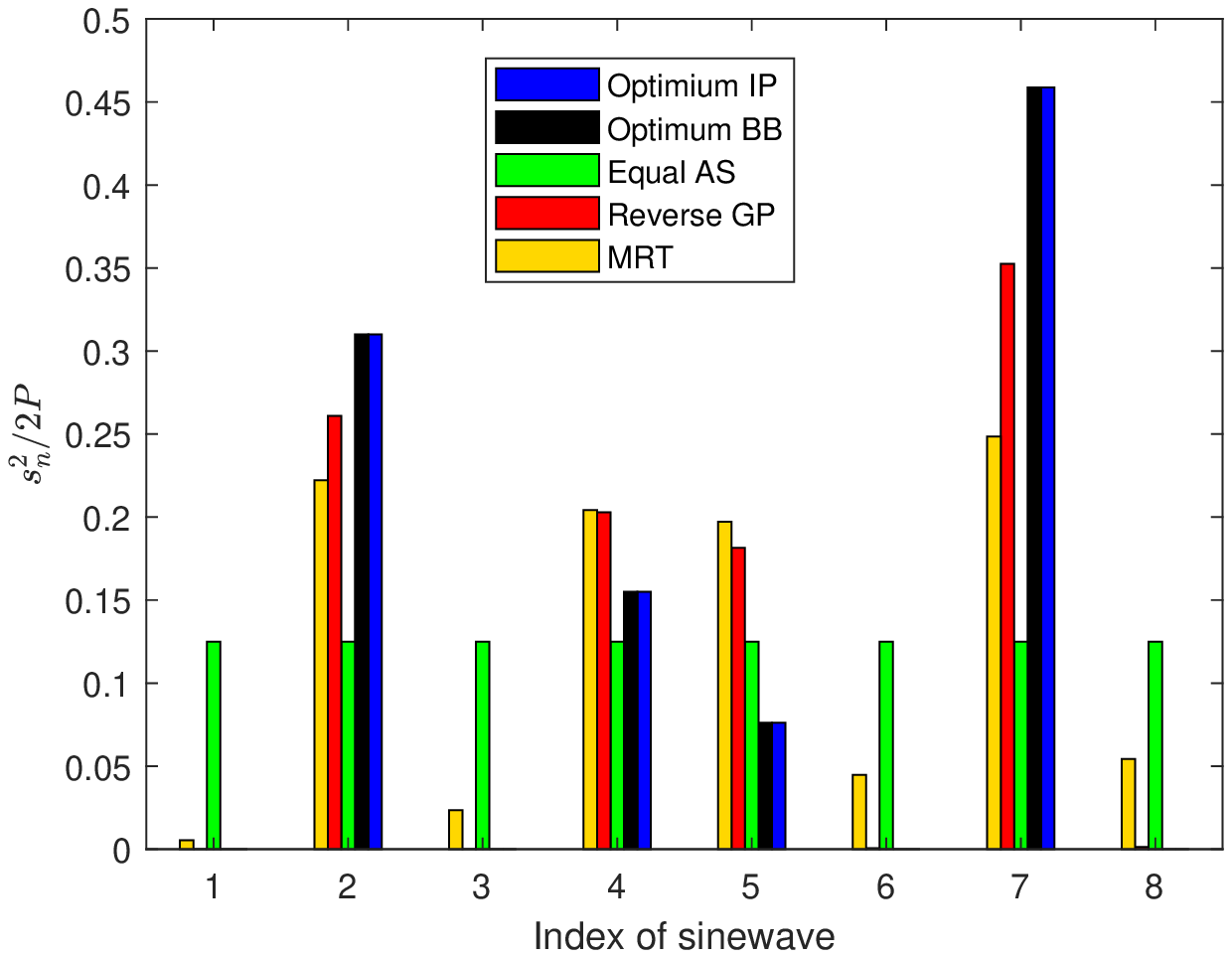}}
		\subfloat[Average output power vs. normalized transmitted power.\label{fig:Output-power-vs. Pin Curvefit}]{\includegraphics[width=3.5in,height=2.5in]{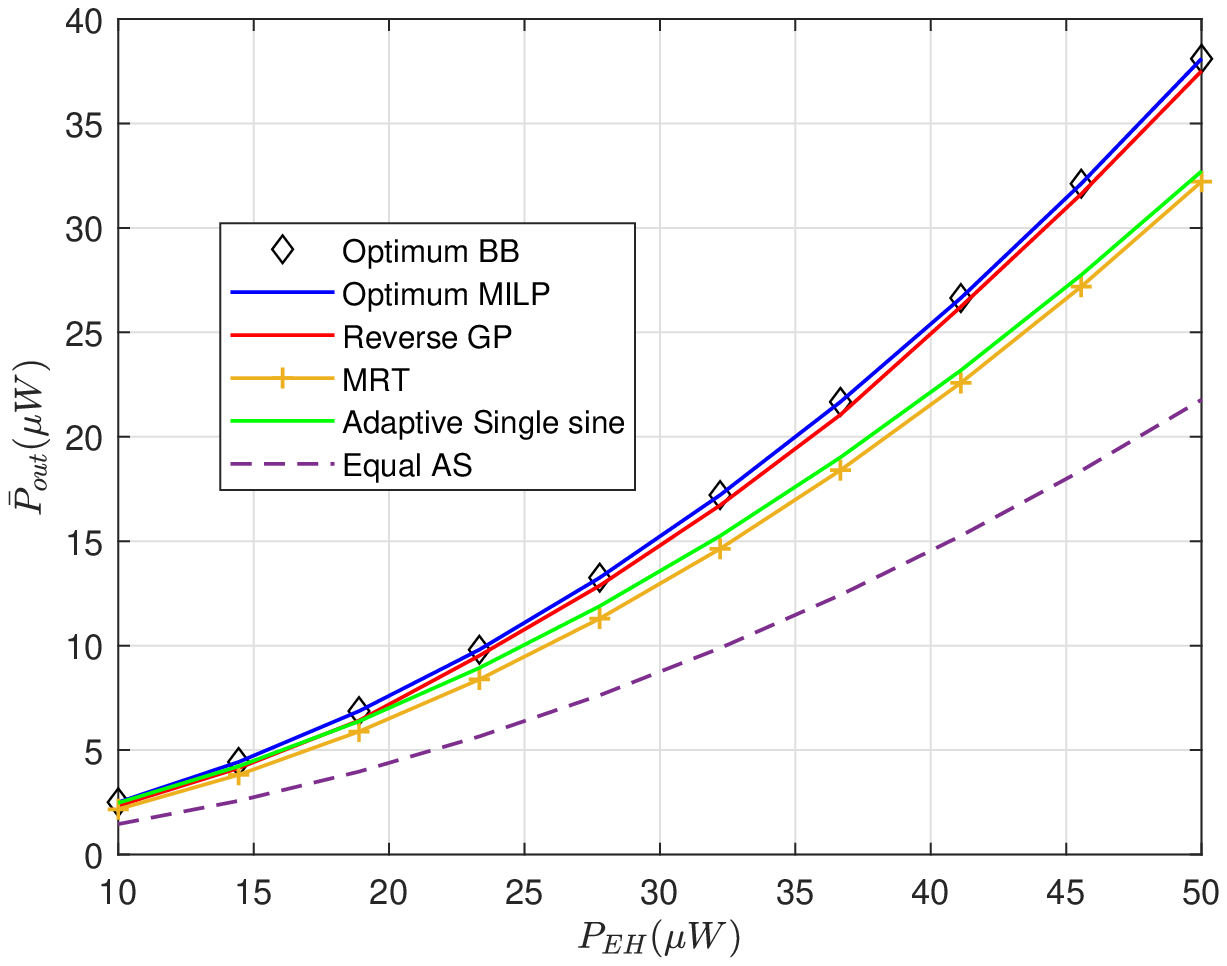}}
		\par\end{centering}
	\caption{Simulation results for the second-order polynomial curve-fitting model.}
		\vspace{-0.2in}
\end{figure}

Second-order polynomial fitting using MATLAB was applied to the simulated
input-output power characteristics as previously shown in Fig. \protect\ref{fig:Curvefit}
to calculate the values of $\beta_{1}$ and $\beta_{2}$ in (\protect\ref{eq:SecondOrderP}).
Using the same channel realization in Fig. \protect\ref{fig:Normalized-channel-frequency diode},
for $M=1$ and $P_{EH}=50\,\mathrm{\mu W}$, the power allocated to
each tone (normalized to the total transmitted power) obtained from
solving the optimization problem resulting from the curve-fitting
model is shown in Fig. \protect\ref{fig:Mutli-tone-amplitudes-Curvefit}.
It can be observed that power is allocated to different tones in a
similar way as in the case of the diode model. The small difference
comes from the fact that $\beta_{1}\neq k_{4}R_{ant}^{2}$ and $\beta_{2}\neq k_{2}R_{ant}$.
The simulated average output DC power is plotted vs. the normalized
transmitted power in Fig. \protect\ref{fig:Output-power-vs. Pin Curvefit}
emphasizes the improvement achieved by optimizing multi-tone excitation.
The results shown in Fig. \protect\ref{fig:Mutli-tone-amplitudes-Curvefit}
and Fig. \protect\ref{fig:Output-power-vs. Pin Curvefit} prove that the second-order
curve-fitting model can be used as an alternative model for the multi-tone
optimization problem.
\subsection{Results for In-Band SWIPT}
\begin{figure}[t]
\begin{centering}
	\subfloat[Normalized channels frequency response (normalized to $h_{max}).$\label{fig:Normalized-channels-frequencySWIPT}]
	{\includegraphics[width=3.5in,height=2.5in]{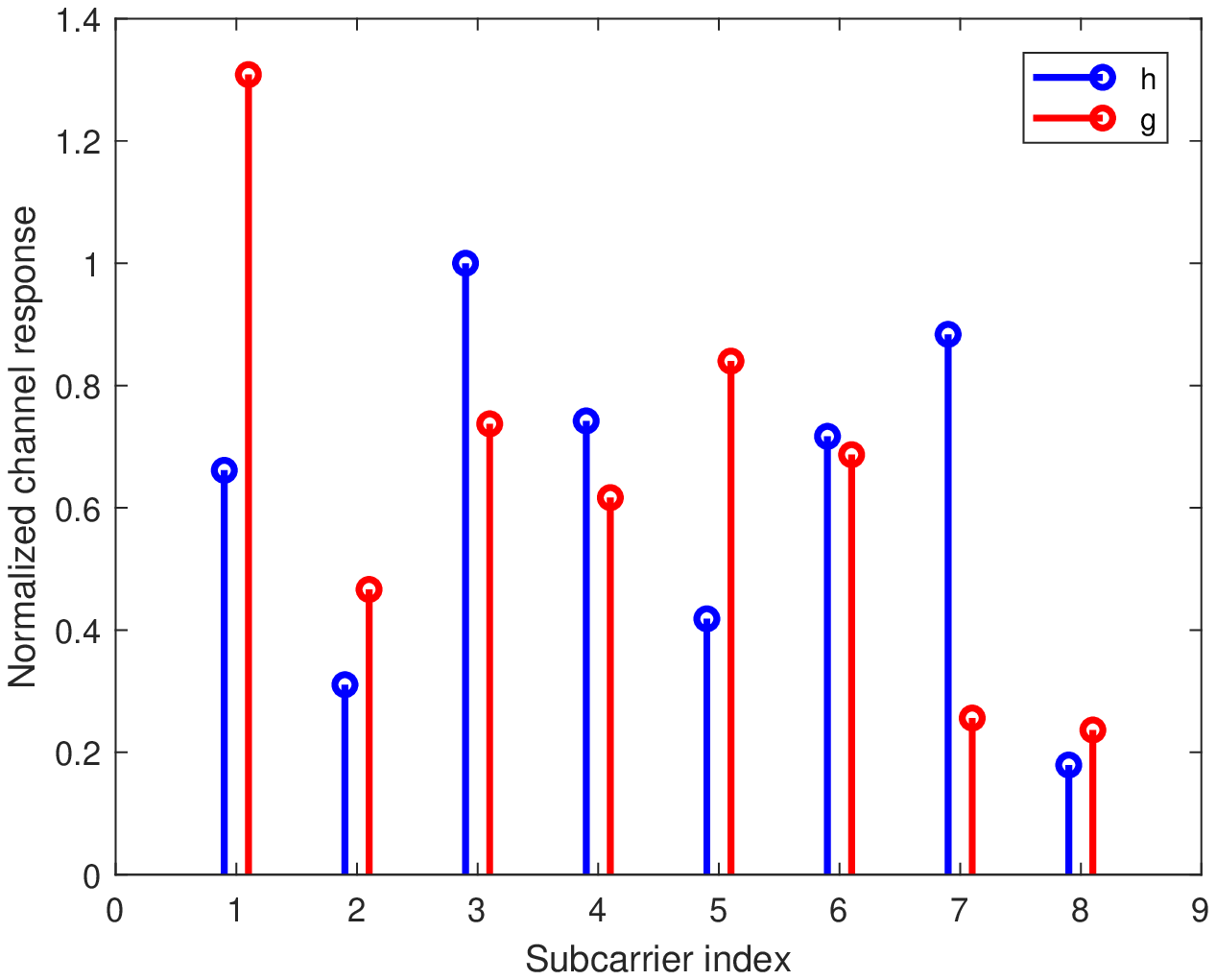}}
	\subfloat[Normalized multi-tone amplitudes\label{fig:Multi-tone-amplitudes-SWIPT}]{
	\includegraphics[width=3.5in,height=2.5in]{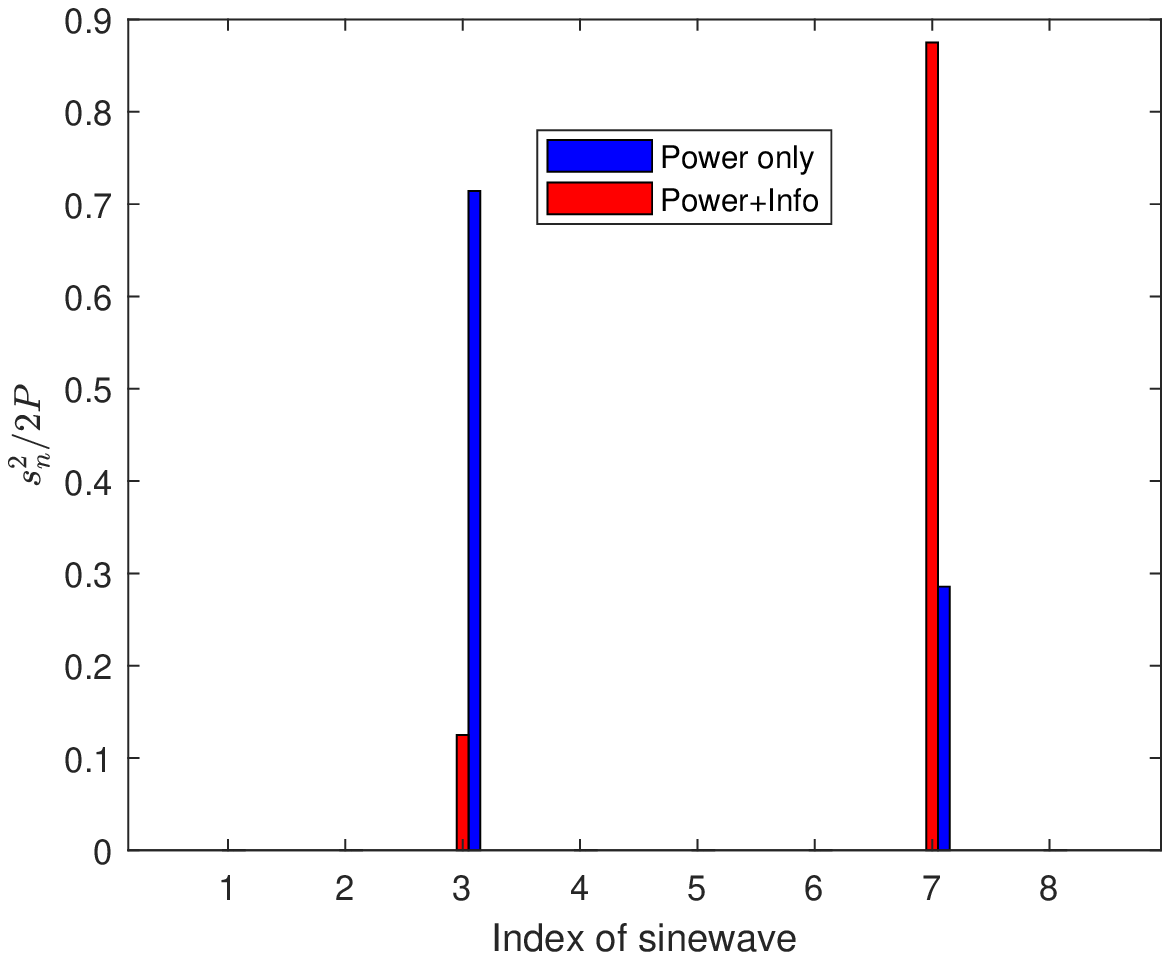}}
\par\end{centering}
\caption{An example for the effect of considering the co-existing information link on power allocation to different tones.}
\vspace{-0.2in}
\end{figure}

\begin{figure}[th]
\begin{centering}
\includegraphics[width=3.5in,height=2.5in]{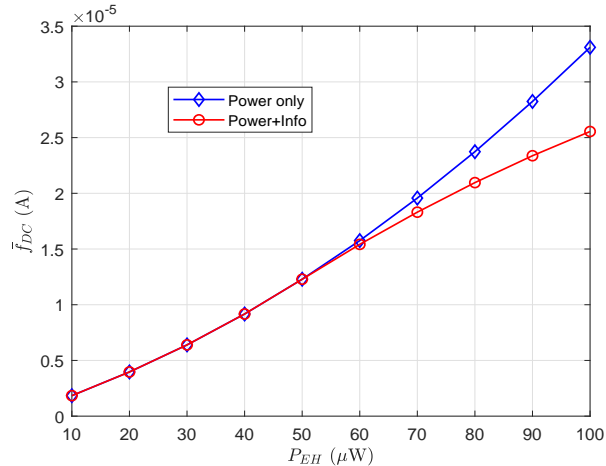}
\par\end{centering}
\caption{Average output current represented by $\bar{f}_{DC}$ vs. normalized
transmitted power given by $P_{EH}$with and without considering the
information link.\label{fig:Average-output-current SWIPT} }
\vspace{-0.2in}
\end{figure}

\begin{figure}[t]
\begin{centering}
	\subfloat[$\bar{f}_{DC}$ vs. the saturation power of the LNA ($P_{sat}.$) \label{fig:Sat Power SWIPT}] {\includegraphics[width=3.5in,height=2.5in]{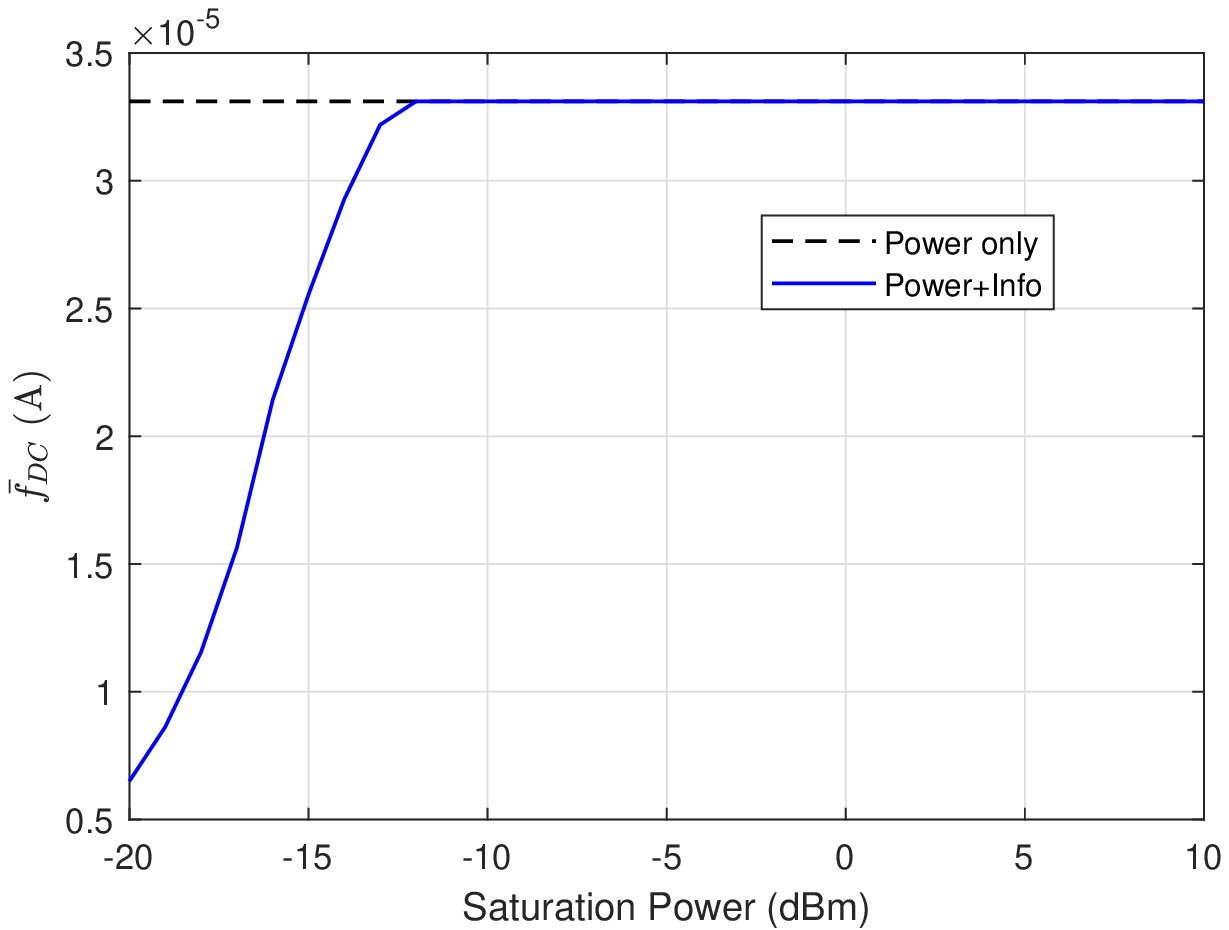}}
	\subfloat[$\bar{f}_{DC}$ vs. the information receiver distance from the power transmitter normalized to the wavelength $\left(\frac{d}{\lambda}\right).$\label{fig:DistanceStudySWIPT}]{\includegraphics[width=3.5in,height=2.5in]{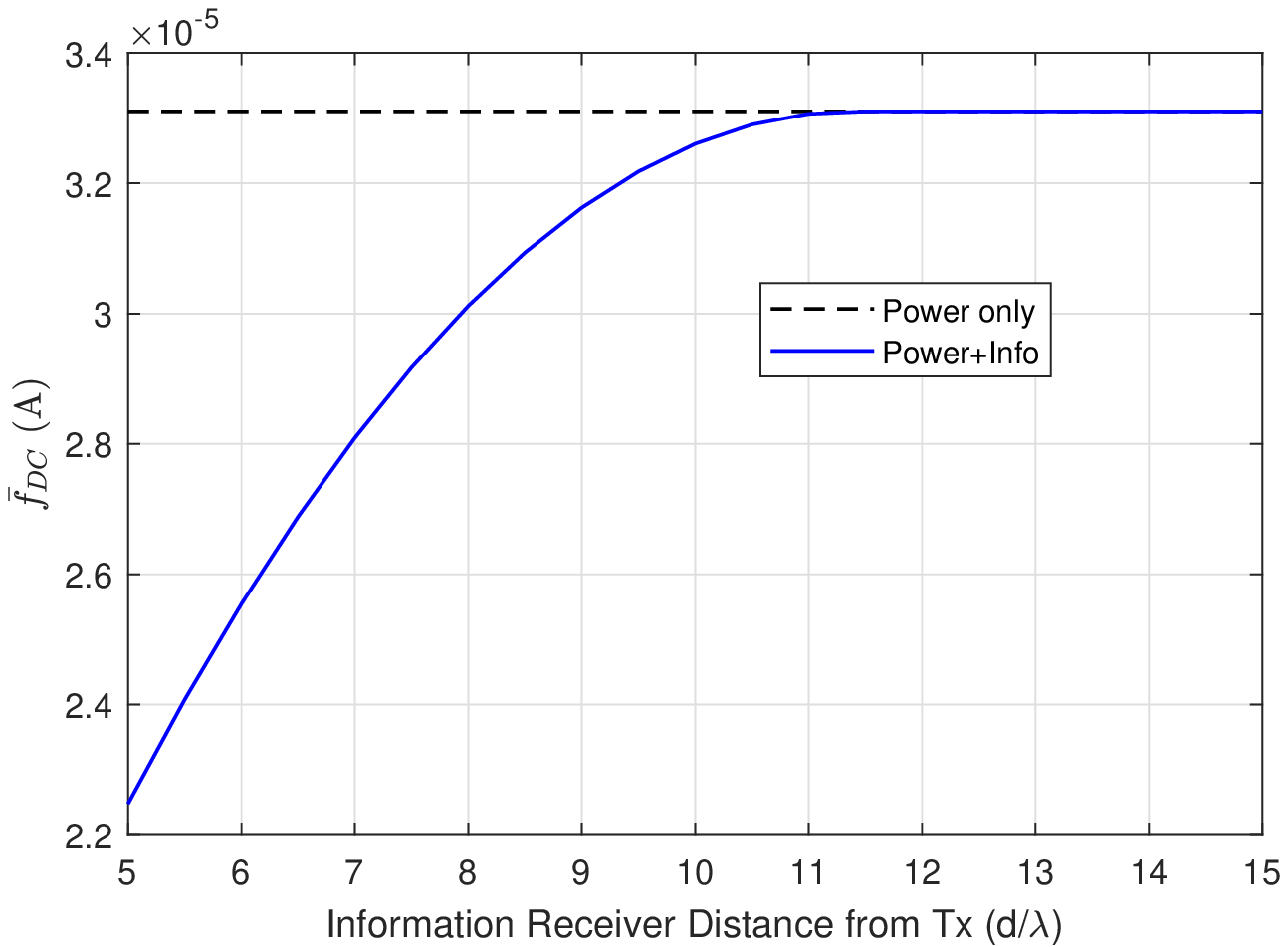}}
\par\end{centering}
\caption{Study of the average output current represented by $\bar{f}_{DC}$ vs. saturation power of the LNA  and the distance of the information receiver from the power transmitter. The energy harvester is fixed at $d=8\lambda$ with $P_{EH}=100\,\mathrm{\mu W}$. The dotted black line indicates the average output current without considering the constraint on $P_{sat}$.} 
\vspace{-0.2in}
\end{figure}

Finally, we consider the co-existing communication link by taking
into account the saturation power of the LNA of the nearby information
receiver. A single antenna is assumed at the power transmitter. The
diode model is adopted for simulations in this subsection. The normalized
transmitted power (maximum possible received power at the energy harvester)
is fixed to $P_{EH}=100\,\mathrm{\mu W}.$ We first assume that the
saturation power $P_{sat}=-15\,\mathrm{dBm}$, and that the nearest
information receiver is located at $d_{2}=7\lambda$, i.e., closer
to the power transmitter from the energy harvester. Accordingly, a
possible realization of the channel coefficients $h$ and $g$ normalized
to $h_{max}$ is plotted in Fig. \protect\ref{fig:Normalized-channels-frequencySWIPT}.
The corresponding optimal power allocation strategy with and without
considering the information link is shown in Fig. \protect\ref{fig:Multi-tone-amplitudes-SWIPT}.
It can be observed from the figure that although sub-carrier\#3 has
a stronger channel to the energy harvester than subcarrier\#7, the
latter has been allocated more power when considering the co-existing
communication link because it has a weaker channel to the information
receiver. By varying the transmitted power, the average output current
is plotted in Fig. \protect\ref{fig:Average-output-current SWIPT}. Taking
the saturation power into consideration affects the maximum possible
output current when increasing the transmitted power. The saturation
power is varied in Fig. \protect\ref{fig:Sat Power SWIPT} and the average
output current is observed. Of course, as the saturation power increases,
the achievable output current increases until it coincides with the
case when having the energy harvester only. In Fig. \protect\ref{fig:DistanceStudySWIPT},
the effect of the distance of the information receiver from the power
transmitter is studied. It can be observed that the effect of the
$P_{sat}$ on the maximum achievable output current decreases as the
distance of the nearest information receiver from the power transmitter
increases. 
\section{Conclusions }
The problem of multi-tone optimization for wireless power transmission
has been studied in this paper. Two non-linear energy harvester models
were discussed: the diode model, and the second-order polynomial curve-fitting
model. We showed that the simple second-order polynomial curve-fitting
model can be used for formulating the multi-tone optimization problem
leading to similar results obtained form the diode non-linear model.
This curve-fitting model can be applied to complicated energy harvester
architectures to simplify the formulation of the multi-tone optimization
problem. We formulated the multi-tone optimization problem as a linear
constrained quadratic program and explained two different solution
methods. We provided a global optimal solution for the tones excitation
coefficients to maximize the output of the energy harvester. We designed
an energy harvester working at $2.4$ GHz for evaluating the performance
of the curve-fitting model with different allocation strategies using
Keysight ADS circuit simulator. Simulation results showed the superiority
of our allocation strategy over all existing power allocation strategies.
The co-existence of information links with wireless power transmission
has been studied in this paper by introducing a constraint on the
received power at the nearby information receiver, which is operating
in the same band as the wireless power transmitter, to prevent its
LNA from saturation and producing harmonics interfering with the information
signal. Performance evaluations highlighted the importance of considering
the saturation power, and the distance of the information receiver
from the power transmitter when designing the multi-tone signal. Considering
co-location of power and information transmission as well as incorporating
non-linear energy harvester models, which are valid in the low-power
region, in the areas of wirelessly powered communications and backscatter
communications is promising for future research. 

\section*{Acknowledgment}
The authors would like to thank Christos I. Kolitsidas for his help
in setting up the experiment described in the introduction of this
paper and for the fruitful discussions during the early stages of
this work. 
\footnotesize
%\bibliographystyle{IEEEtran}
%%%

\end{document}